\documentclass[12pt,preprint]{aastex}

\input epsf
\usepackage{epsfig}
\usepackage{rotating}

\slugcomment{Submitted to Astrophysical Journal}

\shorttitle{Galaxy Clusters in the Swift/BAT era}
\shortauthors{Ajello et al.}

\begin{document}


\title{Galaxy Clusters in the {\it Swift}/BAT era: 
Hard X-rays in the ICM}

\author{M. Ajello\altaffilmark{1},       P. Rebusco\altaffilmark{2},
N. Cappelluti\altaffilmark{1,3},   O. Reimer\altaffilmark{4}, 
H. B\"ohringer\altaffilmark{1},  J. Greiner\altaffilmark{1}, 
N. Gehrels\altaffilmark{5}, J. Tueller\altaffilmark{5} and 
A. Moretti\altaffilmark{6}}
\email{majello@mpe.mpg.de}
\altaffiltext{1}{Max Planck Institut f\"{u}r Extraterrestrische Physik, P.O. Box 1603, 85740, Garching, Germany}

\altaffiltext{2}{Kavli Institute for Astrophysics and Space Research, MIT, Cambridge, MA 02139, USA}

\altaffiltext{3}{University of Maryland, Baltimore County, 1000 Hilltop Circle, Baltimore, MD 
21250}

\altaffiltext{4}{W.W. Hansen Experimental Physics Laboratory \& Kavli Institute
for Particle Astrophysics and Cosmology, Stanford University, USA}

\altaffiltext{5}{Astrophysics Science Division, Mail Code 661, NASA Goddard Space Flight Center, Greenbelt, MD 20771, USA }

\altaffiltext{6}{INAF-OAB, via E. Bianchi 46, Merate (LC) 23807,  Italy}

%
%

\begin{abstract}
We report about the detection of 10 clusters of galaxies in the ongoing
{\it Swift}/BAT all-sky survey. This sample, which comprises mostly
merging clusters, was serendipitously detected
in the 15--55\,keV band. 
We use the BAT sample to investigate the presence of  
excess hard X-rays above the thermal emission. 
The BAT clusters do not show significant 
(e.g. $\geq$2\,$\sigma$) non-thermal hard X-ray emission.
The only exception is represented by Perseus whose high-energy emission
is likely due to NGC 1275.
Using XMM-Newton, Swift/XRT, Chandra and BAT data, 
we are able to produce upper limits on the Inverse
Compton (IC) emission mechanism which are in disagreement with most of
the previously claimed hard X-ray excesses. 
The coupling of the X-ray upper limits on the IC mechanism to radio data
shows that in some clusters the  magnetic field might be larger
 than 0.5\,$\mu$G.
We also derive the first log $N$ - log $S$ and luminosity function
distribution of  galaxy clusters above 15\,keV.
\end{abstract}


%
%

\section{Introduction}
\label{intro}

Galaxy clusters are potentially powerful observational probes of 
dark matter and dark energy. 
However, the use of clusters to measure cosmological parameters becomes 
accessible only when astrophysical uncertainties are well understood and 
controlled. Indeed the non-thermal pressure due to Cosmic Rays (CRs), 
magnetic fields and turbulence, is a source of systematic bias when cluster 
masses are estimated using the assumption of hydrostatical equilibrium  
\citep[e.g.][]{ensslin97}. The detection of clusters' X-ray emission above 
$\sim$20\,keV is a fundamental step towards the  firm grasp of  these 
processes.

It is well understood that clusters of galaxies contain a large amount 
of hot gas, called intracluster medium (ICM), that comprises 10--15\,\% 
of their total mass. 
Already the first X-ray observations indicated the presence of this 
optically thin
 plasma, characterized by an atomic density of about 
$10^{-4}$--$10^{-2}$\,cm$^{-3}$ and temperatures of the order 
of $10^7$--$10^8$\,K 
\citep[e.g.][]{fel66,cat72}.
Also well established is the fact  that the observed X-ray radiation 
from clusters of galaxies is primarily due to the 
thermal bremsstrahlung emission of  such  diffuse hot plasma  
\citep{sar88,petrosian01}. 

However, evidences gathered at different wavelengths 
point to the existence of a non-thermal component.
In particular the detections of an extended synchrotron radio emission 
\citep[e.g.][]{wil70,har78,gio93,gio00,kem01,thierbach03} and, 
more recently, of a possibly non-thermal extreme ultraviolet (EUV) 
excess \citep{lieu96,bow99,bon01,dur02} and soft 
excess \citep[e.g.][]{wer07} suggest the existence of a non-thermal 
X-ray component originating from a population  of relativistic electrons. 
This scenario is confirmed by the detection of 
non-thermal emission in the hard X-ray spectra
of a few galaxy clusters \citep[see e.g.][for a complete review]{kaastra08,rephaeli08}.
Still its actual presence and origin remain controversial 
\citep{renaud06,fus07,wer07,lutovinov08}.

A non-thermal component could arise from a population of point sources 
\citep[e.g. AGN as in][]{katz76,fabian76,fujita07} or from inverse 
Compton (IC) scattering of cosmic microwave background (CMB) photons 
by relativistic electrons \citep[e.g.][]{rep79,sar99}. 
Other possible mechanisms 
are non-thermal bremsstrahlung \citep[e.g.][]{sar99,sar00} and 
synchrotron emission from ultra-relativistic electrons 
\citep[][]{tim04,ino05,eck07}.
If the origin of the high-energy emission is IC scattering,
 then the presence of a large population of relativistic electrons 
(Lorentz factor $>>1000$) is required.
This population could have been accelerated
in shocks of different origin. 
Indeed it could be associated to merger shocks \citep[e.g.][]{fuj03,bru04},
 dark matter bow shocks \citep[e.g.][]{byk00}, ram-pressure stripping of 
infalling galaxies \citep[e.g.][]{dep06}, jets, Active Galactic Nuclei 
outbursts 
\citep[][in the case of radio mini-halos such as in Perseus cluster]{fujita07}, 
accretion shocks \citep[e.g.][]{ino05}. Non-thermal electrons loose energy 
on short timescales (below $1$ Gyr).
 Therefore some models consider a continued supply of primary accelerated
 electrons (i.e. via first order Fermi mechanism), while others assume a 
constant in-situ re-acceleration via CR collisions or second order Fermi 
mechanism.

If clusters are a large reservoir of non-thermal particles, then they 
should emit at higher energies, up to the $\gamma$-rays. Indeed, if CRs 
acceleration is taking place
at the shock fronts then $\gamma$-rays can be produced via IC, non-thermal 
bremsstrahlung and $\pi^0$  decay 
\citep[e.g.][]{rep79,dar95,rei03,reimer04,blasi07}.
A statistical upper limit on the flux above $>100$ MeV was obtained by 
\cite{rei03}, analyzing the emission from $58$ clusters observed with EGRET.

The role of CRs in the formation and evolution of clusters of galaxies
 has been much debated. 
\cite{chu07} suggest that in massive galaxy clusters
hydrostatic equilibrium is satisfied reasonably well, as long as the 
source has not experienced a recent major merger.
However, in non-relaxed clusters the non-thermal pressure due to CRs, 
magnetic fields and micro-turbulence can affect  the mass estimates 
based on hydrostatic equilibrium \citep[e.g.][]{mir95,nag07}. This would 
lead to a 
higher baryonic to total mass ratio.
Knowing the importance of CRs, the mechanisms that heat the ICM and the 
frequency at which it is shocked, is crucial for the upcoming X-ray and 
Sunyaev-Zeldovich surveys \citep[see][]{and07}.

In this paper we report the {\it Swift}/BAT all-sky 
detection of 10 galaxy  clusters in the 15--55\,keV band. 
This constitutes the first complete sample  so far detected at these energies. 
We use this sample to investigate the role of non-thermal processes in clusters.
The structure of the paper is the following.  
In $\S$~\ref{subsec:batsurvey} we describe the 
{\it Swift}/BAT observations 
and discuss the properties of each individual cluster
($\S$~\ref{subsec:individual}). 
In  $\S$~\ref{subsec:non-thermal}, we provide, for all the clusters,
constraints on the non-thermal emission as well as an estimate
of the clusters' magnetic fields ($\S$~\ref{subsec:magnetic}).
The cluster source count distribution 
and the luminosity function are  derived in  $\S$~\ref{sec:pop}.
We discuss  the results of our analysis  in $\S$~\ref{sec:disc}, while
$\S$~\ref{sec:concl} summarizes our findings.

We adopt a Hubble constant of $H_0 = 70$ h$_{70}$\,km s$^{-1}$ Mpc$^{-1}$, 
$\Omega_M$ = 0.3 and $\Omega_\Lambda$ = 0.7.
Unless otherwise stated errors are quoted at the 90\,\% confidence
level (CL) for one interesting parameter and solar abundances
are determined using the meteoritic values provided in \cite{anders89}.

%
%
\section{The BAT X-ray Survey}
\label{subsec:batsurvey}
The Burst Alert Telescope  \citep[BAT;][]{barthelmy05}, on board the 
Swift satellite \citep{gehrels04},
 represents
a major improvement in sensitivity for imaging of the hard X-ray sky.
BAT is a coded mask telescope with a wide field of view 
(FOV, $120^{\circ}  \times 90^{\circ}$ partially coded) aperture
sensitive in the 15--200\,keV domain.
BAT's  main purpose is to locate Gamma-Ray Bursts (GRBs).
While  chasing new GRBs, 
BAT surveys the hard X-ray sky  with an unprecedented sensitivity.
Thanks to its wide FOV  and its  pointing strategy, 
BAT monitors continuously up to 80\% of the sky every day.
Results of the BAT survey \citep{markwardt05,ajello07a} 
show that BAT  reaches a sensitivity of $\sim$1\,mCrab in 1\,Ms of exposure.
Given its sensitivity and the large exposure already accumulated in the whole
sky, BAT poses itself as an excellent instrument for looking 
for the (faint) emission of galaxy clusters above 15\,keV.
\\\\
For the analysis presented here, we used all the available data
taken from January 2005 to March 2007.
Since most of the cluster emission is expected to be thermal and thus
rather soft, the survey chosen energy interval is 15-55\,keV. The lower
limit is dictated by the energy threshold of the detectors. The upper limit
was chosen as to avoid the presence of strong background lines which
could worsen the overall sensitivity. 
The data screening was performed according to \cite{ajello07a}.
The all-sky image is obtained as  
the weighted average of all the shorter observations.
The average exposure time in our image is 3\,Ms, being 1.3\,Ms and 5\,Ms the minimum
and maximum exposure times respectively. 
The final image shows a Gaussian normal noise.
Source candidates were  identified 
as excesses above  the 5\,$\sigma$ level.
All these objects are then fit with the BAT point spread function (using the standard BAT
tool {\it batcelldetect}) to derive the best source position.
\\\\
As shown in \cite{ajello07a} cross-correlating the BAT sources with 
the ROSAT All-Sky Survey Bright Source Catalogue \citep{voges99}
provides an easy and solid way to identify a large fraction ($\sim$70\%) of them.
Most of the uncorrelated sources are not present in the ROSAT survey
because of absorption (either along the line of sight or intrinsic to the source).
The unidentified sources are  targeted by Swift/XRT  which
in less than 10\,ks can pinpoint the exact counterpart 
\citep[e.g.][]{tueller05a,tueller05b,kennea05}.
\\
The details about the complete source list will be given in an
upcoming publication.
Here we report about the detection of galaxy clusters above 5\,$\sigma$
in the 15--55\,keV band.

%
%
\subsection{Clusters Identification}
Identifying clusters of galaxies as counterpart of BAT objects is not a 
straightforward process. Indeed, coded mask telescopes are rather insensitive
to diffuse sources which extend over angles much larger than the projection
of the mask tile on the sky (i.e. a few tens of  arcmin for BAT).
Even though procedures exist to quantify the extent of diffuse sources in
coded mask instruments 
\citep[see][for the case of the Coma Cluster]{renaud06,renaud06a,lutovinov08},
their application is limited only to high signal-to-noise (S/N) objects.
Given the extent of the BAT PSF (22\arcmin\ ), Coma is the only
object whose emission is clearly extended in our investigation.
Thus, for all other objects,
the morphology of the source
cannot be used to understand whether the BAT source is associated to the cluster
or only to its brightest AGN. We therefore performed a spectral analysis
(see \S~\ref{subsec:individual}) of those 
BAT sources that are spatially associated with galaxy clusters.
All sources presented here  show a significant
thermal component that we interpret as thermal bremsstrahlung from the ICM
and thus are securely associated with the proposed clusters.
Our sample contains 10 galaxy clusters.
Table~\ref{tab:clusters} reports the position,
significance, total exposure time and other details 
of all the detected clusters.

%
%
\subsection{Spectral Analysis}
\label{subsec:individual}

For each galaxy cluster we extracted a 15--195\,keV spectrum  with the method 
described in \cite{ajello07b}. Here we recall the main steps: the details can be
found in the aforementioned paper. For a given
source, we extract a spectrum from each observation where the source is in the
field of view. These spectra are corrected for residual background contamination
and for vignetting; the 
per-pointing spectra are then (weighted) averaged to produce the final source
spectrum. Thus, the final spectrum represents the average source emission 
over the time-span considered here (2.5 years). The accuracy of these
spectra is discussed in $\S$~\ref{sec:accuracy}.

For all the clusters, we extracted a 0.3--10\,keV spectrum  using
archival observations of XMM-Newton, Chandra, and Swift/XRT.
Considering that for BAT all clusters, except Coma, are point-like objects
we extracted (unless otherwise stated)
all cluster photons  within 10\arcmin\ from the position of the
BAT centroid. In most cases, this selection allows us to include most of  the
emission of the cluster. For those cases where there is clearly emission
outside of our selection region\footnotemark{},
\footnotetext{In some cases the extent of the 
selection region is limited by the size of the CCD.} 
we accounted for the missed flux
using the beta profiles available in  literature. The 
details are given in the case-by-case section 
($\S$~\ref{sec:specanalysis}).
The level of the
background was evaluated in regions of the CCDs not contaminated by
the cluster emission or using blank-sky observations 
\citep[e.g][]{lumb02,read03}. In all cases, we 
considered the systematic
uncertainty connected to the background subtraction, in the 0.3--10\,keV band,
to be 2\,\%.
All spectra were rebinned in order to have
a minimum of 50 counts ($\geq$7\,$\sigma$) per bin.

As a standard procedure, we started fitting all the spectra
with the most simple and plausible spectral model.
In all cases this was a single-temperature thermal model
with absorption fixed at  the Galactic value.
Only if the value of the $\chi^2/dof$ was greater than 1, we tried
to add a second thermal model or a power law. In this case
we chose the model which produced the best improvement in the fit
(evaluated using the F-test) and the best residuals.

Various authors have reported detections of hard X-ray excesses
for some of the clusters present in our sample. For those
cases where we do not detect directly such component, we tested
whether our data are consistent with  the reported non-thermal hard
X-ray emission.
This was done adding a power law to the thermal model used. 
We fixed the power-law
index to 2.0, which is a value generally accepted for the non-thermal
hard X-ray
component generated by IC of relativistic  electrons off CMB photons
\cite[e.g.][]{reimer04,nevalainen04}. 
We then let the power-law normalization vary until the $\Delta \chi^2$
increment was larger than  2.7(6.64). 
According to \cite{avni76}, this gives the 90\,\% (99\,\%) confidence
level on the parameter of interest. This allows us to investigate
the level of non-thermal flux which is consistent with our data.

\subsection{Accuracy of BAT spectra}
\label{sec:accuracy}
\begin{figure*} 
    \begin{center}
     \begin{tabular}{cc}
   \includegraphics[scale=0.45,angle=0]{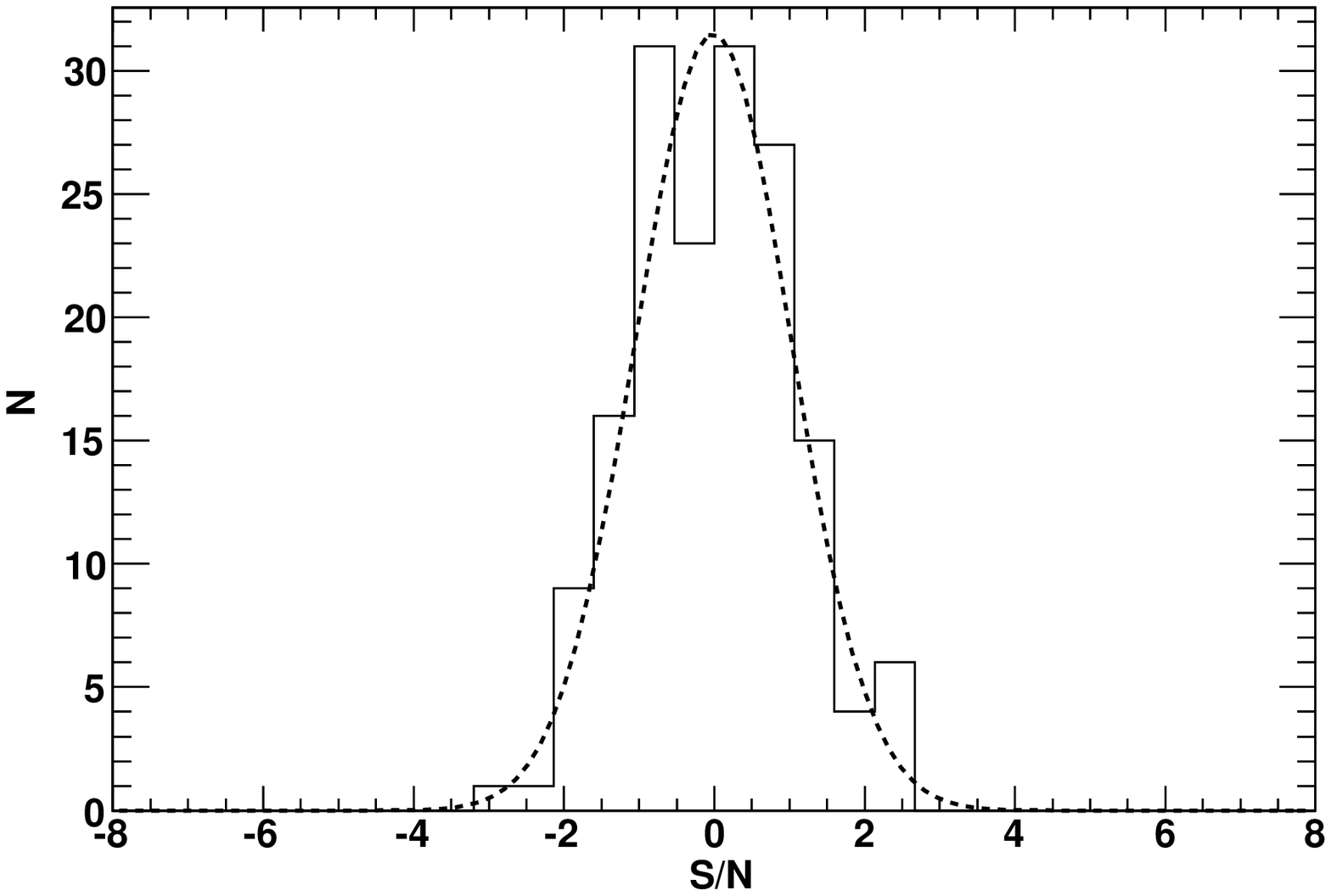} &
   \includegraphics[scale=0.45,angle=0]{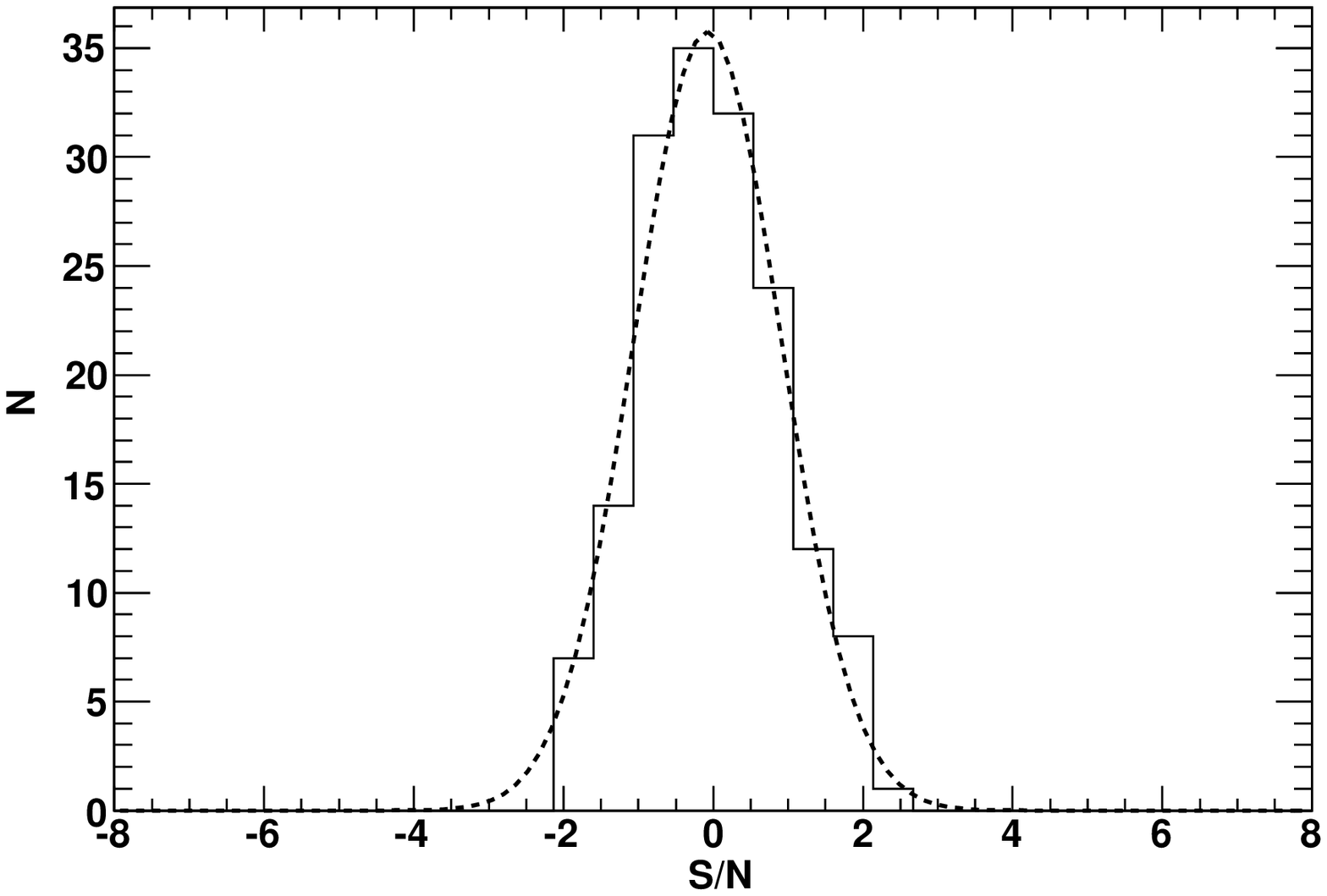} \\
    \end{tabular}
    \end{center}
    \null\vspace{-7mm}
   \caption{Assessment of systematic errors for two representative
energy channels: 18--22\,keV (left) and 57.6--75.4\,keV (right). 
The histograms show the distribution of S/N for 160 random positions (noise)
in the sky away from known or detected sources. The dashed line is a fit 
to the data using a Gaussian profile. The 1\,$\sigma$ 
widths of the Gaussian profiles are compatible with 1.0.}
  \label{fig:syst}
\end{figure*}

When dealing with  spectral features which are at the limiting sensitivity
of a given instrument, it is important to make sure that all systematic
uncertainties have been carefully taken care of. In order to test the
reliability of our spectral extraction method, we extracted $>$160 spectra
at random positions in the sky at least 30\arcmin\
away from the potential (or detected) X--ray sources reported in
the INTEGRAL reference catalog \citep{ebisawa03}. The mean (raw) exposure
of our spectral sample is 4.6\,Ms. In each energy channel, 
the average flux is consistent within 1\,$\sigma$ with zero as expected
for pure noise and for efficient background subtraction.
Moreover, the S/N  distributions  (i.e. flux divided by its error 
in a given
energy channel; examples are shown in Fig.~\ref{fig:syst})
are all consistent with Normal Gaussian distributions. 
Both findings show that our spectra are trustable in the whole energy range
(15--200\,keV) and that uncertainties are well estimated.

\begin{figure}
\begin{center}
\includegraphics[scale=0.6]{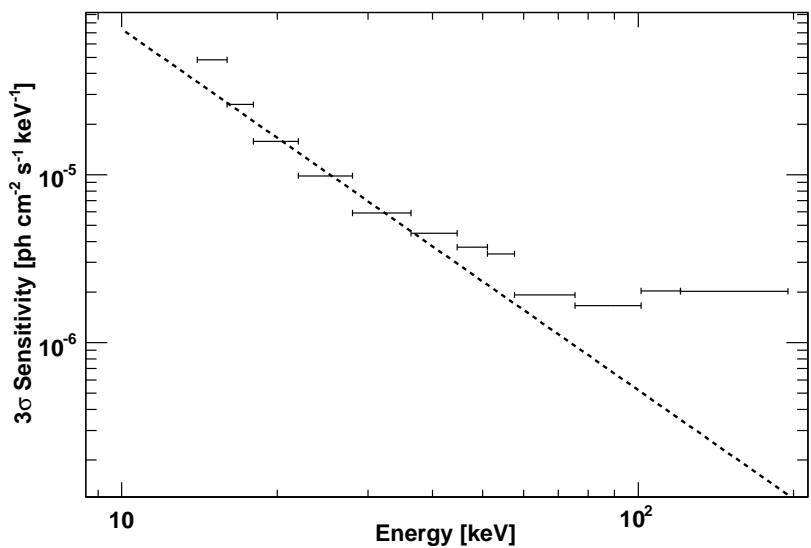}
\caption{3\,$\sigma$ average 
spectral sensitivity as a function of energy based
on the analysis of 160 randomly extracted spectra. The dashed line
is the Crab Nebula spectrum divided by 1000.
\label{fig:sens}}
\end{center}
\end{figure}

Moreover, we can use the randomly extracted spectra to measure the average
spectral sensitivity of BAT in a given energy channel. 
This is done deriving for each energy
channel the standard deviation of the flux distribution. As shown in
Fig.~\ref{fig:sens}, the 3\,$\sigma$ sensitivity in each energy channel
is very close, except above 100\,keV, to 1\,mCrab.

\subsection{Individual Cluster Analysis}
\label{sec:specanalysis}

\subsubsection{Perseus}
\begin{figure*} 
    \begin{center}
     \begin{tabular}{cc}
    \includegraphics[scale=0.4,angle=0]{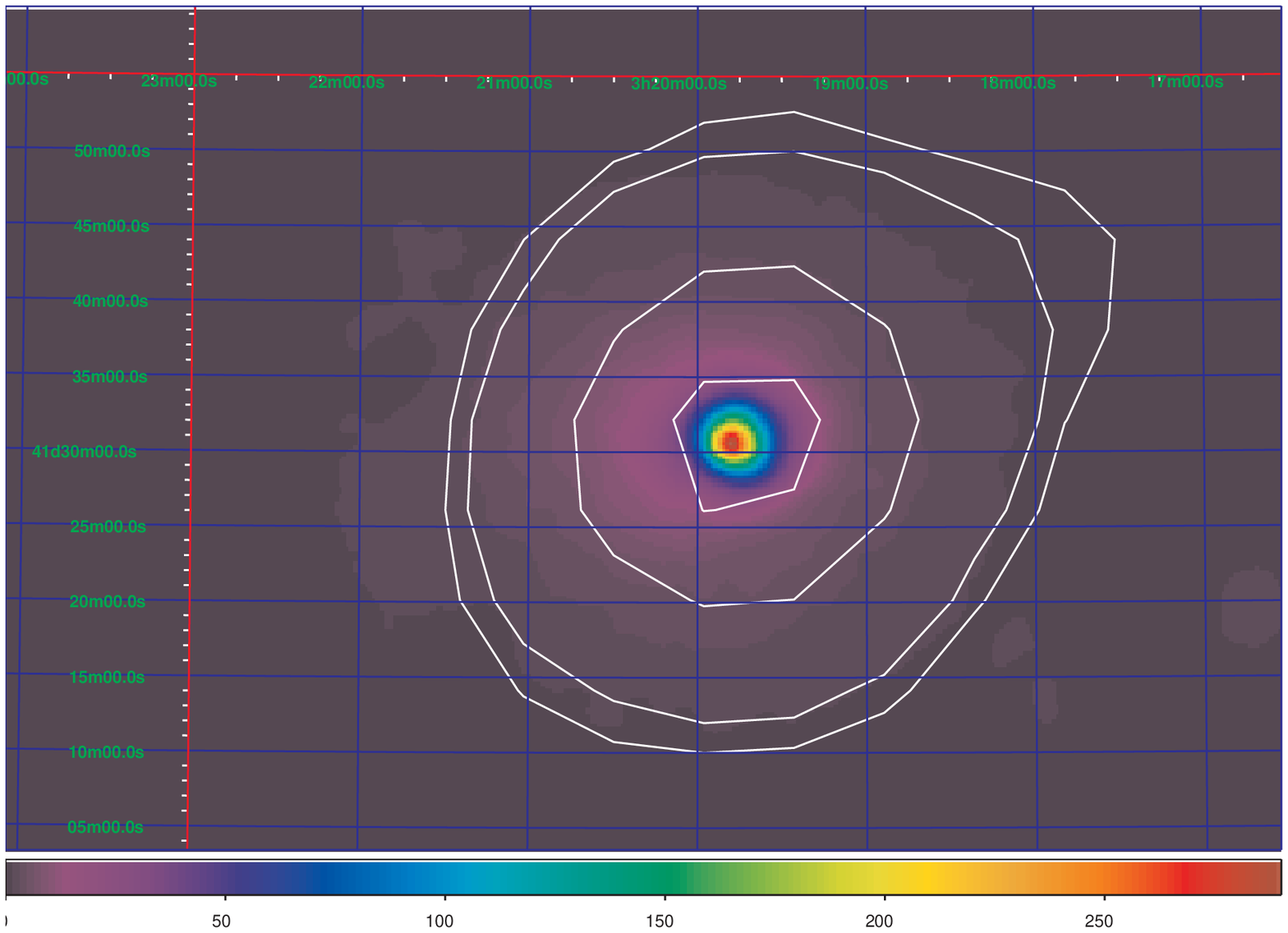} &
   \includegraphics[scale=0.45,angle=0]{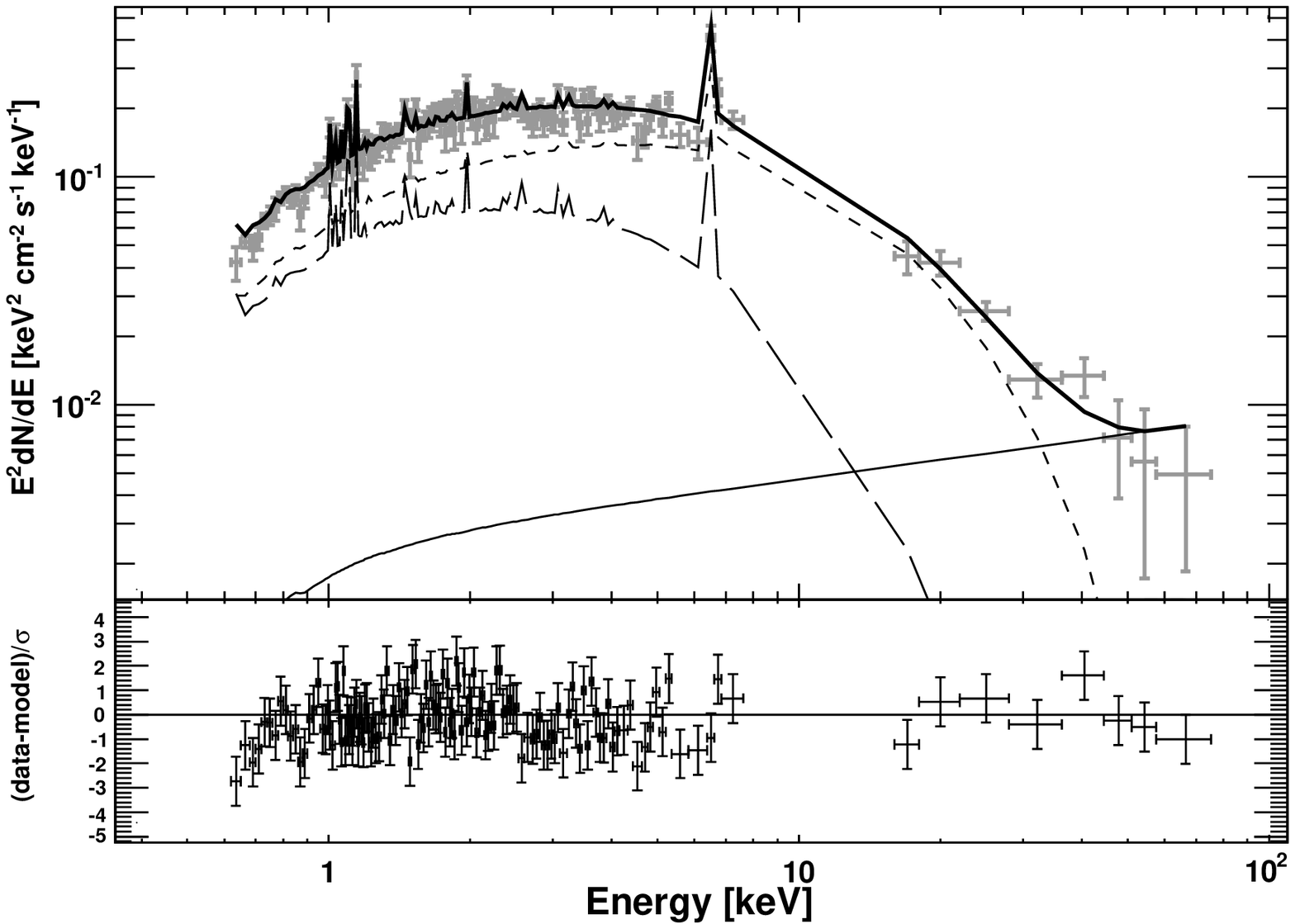} \\
    \end{tabular}
    \end{center}
    \null\vspace{-7mm}
   \caption{
{\it Left Panel:}  ROSAT 0.1--2.4\,keV surface brightness of Perseus
with BAT significance contours superimposed. The contours range from 
2.5\,$\sigma$ to 28\,$\sigma$.
{\it Right Panel:} Joint XRT--BAT spectrum of Perseus. The best fit 
(thick solid line) is the sum of two thermal models 
(dashed and long-dashed line) and of a power-law component (thin solid line).
}
  \label{fig:perseus}
\end{figure*}

{\bf Swift J0319.8+4130} is certainly associated with the Perseus cluster (Abell 426).
The BAT detection (see Fig.~\ref{fig:perseus}) is well centered on the cluster.
Perseus is one of the most studied galaxy clusters
and its detection in X-rays dates back to the seventies \citep{fritz71,for72}.
XMM-Newton observations \citep{churazov2003} 
showed that the central region is contaminated by 
the emission of the AGN hosted by the brightest galaxy in Perseus, NGC 1275.
A hard X-ray component has been detected with HEAO 1 by \cite{pri81}. 
\cite{nevalainen04} used BeppoSAX and previous RXTE measurements 
to prove that this non-thermal component is variable and must 
therefore be connected to the central bright AGN.
\cite{san07} reported, using Chandra,  the presence of 
non-thermal X-ray emission 
in the core of Perseus in correspondence of the radio mini-halo 
\citep[i.e.][]{gis79,git02}. This non-thermal emission, which
displays a power-law behavior with photon index of 2.0, seems to exceed the
flux of NGC 1275 by a factor $\sim$3 \citep{sanders04}. 

The BAT spectrum shows evidences of an hard X-ray excess.
Indeed, it can be fit by a steep power-law (photon index
of 3.5$\pm 0.1$ and $\chi^2_{red}=2.3$) while it rejects
a simple bremsstrahlung fit ($\chi^2_{red}=3.6$).
The fit improves ($\chi^2_{red}=1.50$) if we use a composite
model, sum of the  (bremsstrahlung like) gas emission and the (power-law
like) AGN emission.  The improvement of the fit is statistically significant
as confirmed by the f-test probability of 1.2$\times10^{-2}$.
The best fit temperature is 6.4$^{+2.3}_{-2.3}$\,keV
and the photon index is 2.5$^{+1.9}_{-1.0}$.
If we fix the photon index  at the value (1.65) determined
by \cite{churazov2003}, we derive an extrapolated 0.5--8.0\,keV
luminosity of $\sim0.4 \times10 ^{42}$\,erg s$^{-1}$ which 
is in agreement with the luminosity measured by 
XMM-Newton. This supports the idea that the hard-tail seen in the BAT
spectrum is due to NGC 1275 and not to a non-thermal component originated
in the ICM. Moreover, if we extrapolate, using a
power-law with photon index of 2.0, the non-thermal
flux found in the 2--10\,keV range by \cite{san05} to the 50--100\,keV
band we get a value of 2.7$\times 10^{-11}$\,erg cm$^{-2}$ s$^{-1}$. 
This flux is a factor $\sim$4 larger than the total cluster flux observed by
BAT in the same energy band. Recently \cite{molendi08}  analyzed
a long XMM-Newton observation and did not find evidence for non-thermal
emission. According to them the discrepancy between the Chandra and
XMM-Newton results is due to a problem in the effective area calibration
of Chandra.

An XRT observation of 5.4\,ks was carried out in July 2007.
Given the size of the XRT CCD, we extracted all source photons
within 6\arcmin\ from the BAT centroid. 
The surface brightness profile of Perseus is best described
by the sum of a power-law and of a beta model.
Adopting this model, as suggested by \cite{ettori98}, yields that $\sim$94\,\%
of the total cluster emission falls within our selection. 
The joint XRT--BAT spectrum can be fitted by a sum of two
APEC \citep{smith01} models and a  power law. 
The low-temperature component, which accounts for the cool core of the 
cluster, has a temperature of 3.0$^{+0.4}_{-0.7}$\,keV
and an abundance of 0.43$^{+0.20}_{-0.16}$ solar. The warmer component
displays a temperature of 6.40$^{+0.62}_{-0.71}$\,keV and an abundance
of 0.31$^{+0.15}_{-0.15}$ solar. These results are in line with the
analyses of  \cite{churazov2003} and \cite{san05}. 
Both the power-law photon index of 1.7$^{+0.3}_{-0.7}$ and the 
the luminosity in the 0.5--8.0\,keV band of $\sim8 \times 10^{-42}$\,erg s$^{-1}$
are compatible with the values found, for NGC 1275, by  \cite{churazov2003}
and the ones determined in the next section. 
The photon index is slightly harder than the average photon index
(2.0) of BAT AGN, however it is not unusual for radio-loud objects
\citep[e.g.][]{ajello07b}.

\subsubsubsection{The Nucleus of Perseus}
In order to study more in details the nuclear emission,
 we analyzed a 125\,ks long
XMM-Newton observation (observation 0305780101).
We extracted the spectrum of the nucleus in a radius of 25\arcsec\
and evaluated the local background in an annulus around the source region.
We note that  the results presented here are not sensitive to the radius
of the extraction region if this is in the 10\arcsec\--30\arcsec\  range.
The 0.2--9.0\,keV  spectrum of the nucleus 
is well fitted ($\chi^2$/dof=960.1/731) by an absorbed power-law model with 
absorption consistent with the Galactic one and a photon index
of 1.60$\pm 0.02$. Moreover, we find evidence (at the 95\,\% CL) of 
a K$\alpha$ Iron line with equivalent width of 90.2$\pm45.0$\,eV.
An absorbed APEC model with a temperature of 12.6$\pm0.7$\,keV
provides a worst fit ($\chi^2$/dof=1167.1/732) to the data.
In particular, the absorption  would be required to be lower
than the Galactic one at 99\,\% CL. This fact, in conjunction with
the presence of the Iron line, supports the evidence that the non-thermal
emission in the nucleus of the Perseus cluster is produced
by the central AGN.
The non-thermal luminosities in the 0.5--8.0\,keV and 2.0--10.0\,keV
bands are 7.6$^{+0.2}_{-0.2}\times 10^{42}$\,erg s$^{-1}$ and
  6.5$^{+0.2}_{-0.2} \times 10^{42}$\,erg s$^{-1}$ respectively. 
In order to check these results we extracted a similar spectrum
of the nucleus using {\em Swift}/XRT data and selecting an extraction
region of 10\arcsec\ . The XRT data are compatible with the 
XMM-Newton one. Indeed, fixing the absorption at the Galactic value
we find that the XRT data are compatible with a power-law
model with a photon index of 1.6$\pm0.1$ and that the 
2.0--10.0\,keV luminosity is 8.2$^{+1.1}_{-1.0} \times 10^{42}$\,erg s$^{-1}$.
Thus, the nucleus displays a moderate variability
between the  XMM-Newton and {\em Swift}/XRT observation epochs.
This supports, once more, the interpretation that the non-thermal
emission is produced by the central AGN.

\subsubsection{Abell 3266}
\begin{figure*} 
    \begin{center}
     \begin{tabular}{cc}
    \includegraphics[scale=0.4,angle=0]{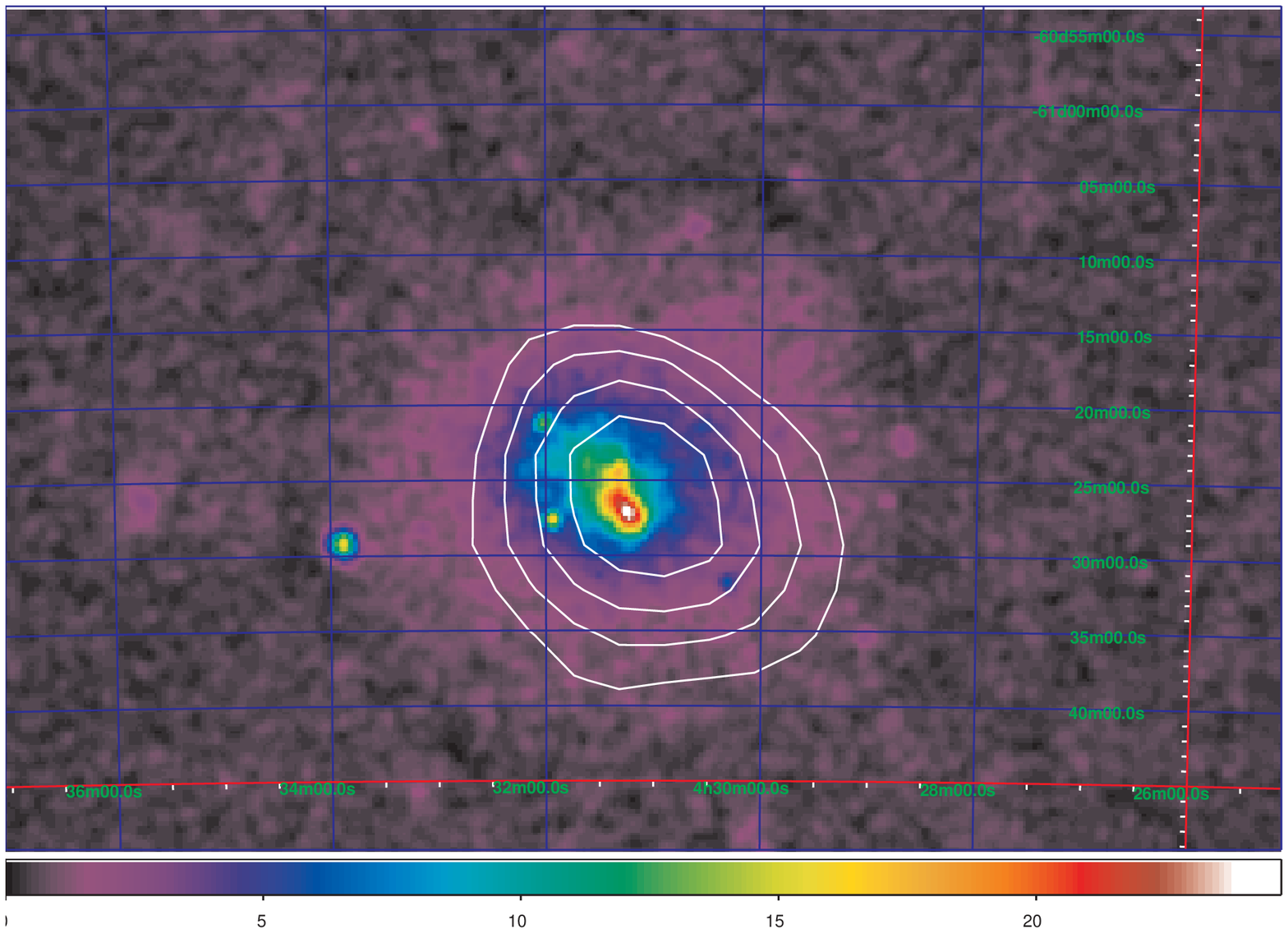} &
   \includegraphics[scale=0.45,angle=0]{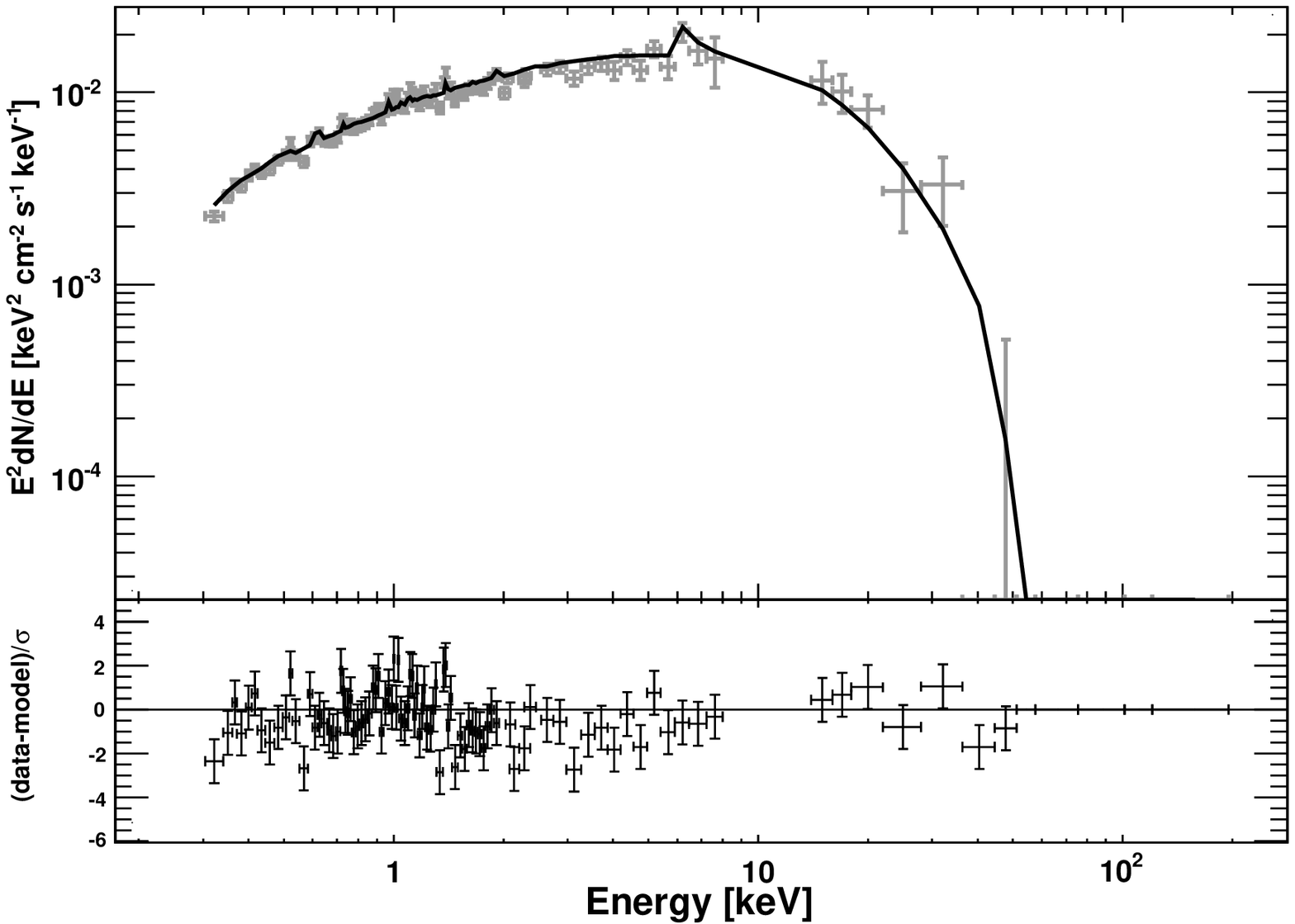} \\
    \end{tabular}
    \end{center}
    \null\vspace{-7mm}
   \caption{
{\it Left Panel:} ROSAT 0.1--2.4\,keV surface brightness of Abell 3266
with BAT significance contours superimposed. The contours range from 
2.5\,$\sigma$ to 5.5\,$\sigma$.
{\it Right Panel:}
Joint fit to XMM-Newton--BAT data for Abell 3266 with a thermal model.
The best  model is shown as solid line.
}
  \label{fig:abell3266}
\end{figure*}

{\bf Swift J0431.3-6126} is associated with Abell 3266. 
Figure~\ref{fig:abell3266} shows that the BAT source is well centered
on the cluster emission as seen by ROSAT.
A 3266
(also known as Sersic 40-6) was first detected in X-rays by the UHURU satellite
 \citep{giacconi1972}. Accordingly to many authors \citep[e.g.][and references
therein]{sauvageot2005,fin06} Abell 3266 recently underwent a major merger, 
probably with a subcluster that was stripped during the encounter with A 3266 
dense core.
\cite{degrandi1999} and \cite{nevalainen04} observed Abell 3266 with BeppoSAX.
 The first group modeled the BeppoSAX broad-band
spectrum (2--50\,keV) with a simple
optically thin thermal emission model at the temperature of 8.1$\pm 0.2$\,keV,
 while the second group found a marginal evidence (0.8\,$\sigma$)
of non-thermal X-ray excess.

The BAT spectrum, shown in Fig.~\ref{fig:abell3266}, 
is consistent with the findings of \cite{degrandi1999}.
 A bremsstrahlung
model with a plasma temperature of  6.9$^{+2.5}_{-1.8}$\,keV 
provides indeed a good fit to the data ($\chi^2/dof = 7.2/10$).
XMM-Newton observed Abell 3266 for 8.6\,ks in September, 2000.
The cluster is not centered on the EPIC-PN CCD. Thus we could extract
only photons within a circular region of $\sim$8\arcmin\ radius
centered on the BAT centroid. In order to estimate the flux
missed by our selection, we adopt, for the cluster surface brightness,
a beta profile 
with $\beta$=0.51 and core radius R$_c$=3.1\arcmin\ \citep{sauvageot2005}.
According to our estimate 80\,\% of the total cluster flux is contained
in our selection. Therefore, when fitting jointly the XMM-Newton and the 
BAT data, we use such cross-normalization factor.
The combined XMM-Newton--BAT spectrum is well fitted by a single APEC model
with a plasma temperature of 8.0$^{+0.4}_{-0.4}$\,keV and 
0.41$^{+0.13}_{-0.13}$ solar abundance.
We derive a 99\,\% CL limit on the non-thermal 50--100\,keV flux of
5.70$\times10^{-13}$\,erg cm$^{-2}$ s$^{-1}$.

Extended radio emission correlated with A 3266 has been reported 
\citep{rob90,bro91}. In order to estimate the magnetic field 
(see $\S$~\ref{subsec:non-thermal}, Table 3), 
we adopt the radio data from \cite{bro91}, based on the Parkes catalogue, 
namely a  flux density  
$S_{2700~ \textrm{MHz}}= 1.070$\, Jy and a spectral index $\alpha=0.95$.

\subsubsection{Abell 0754}
\begin{figure*} 
    \begin{center}
     \begin{tabular}{cc}
    \includegraphics[scale=0.4,angle=0]{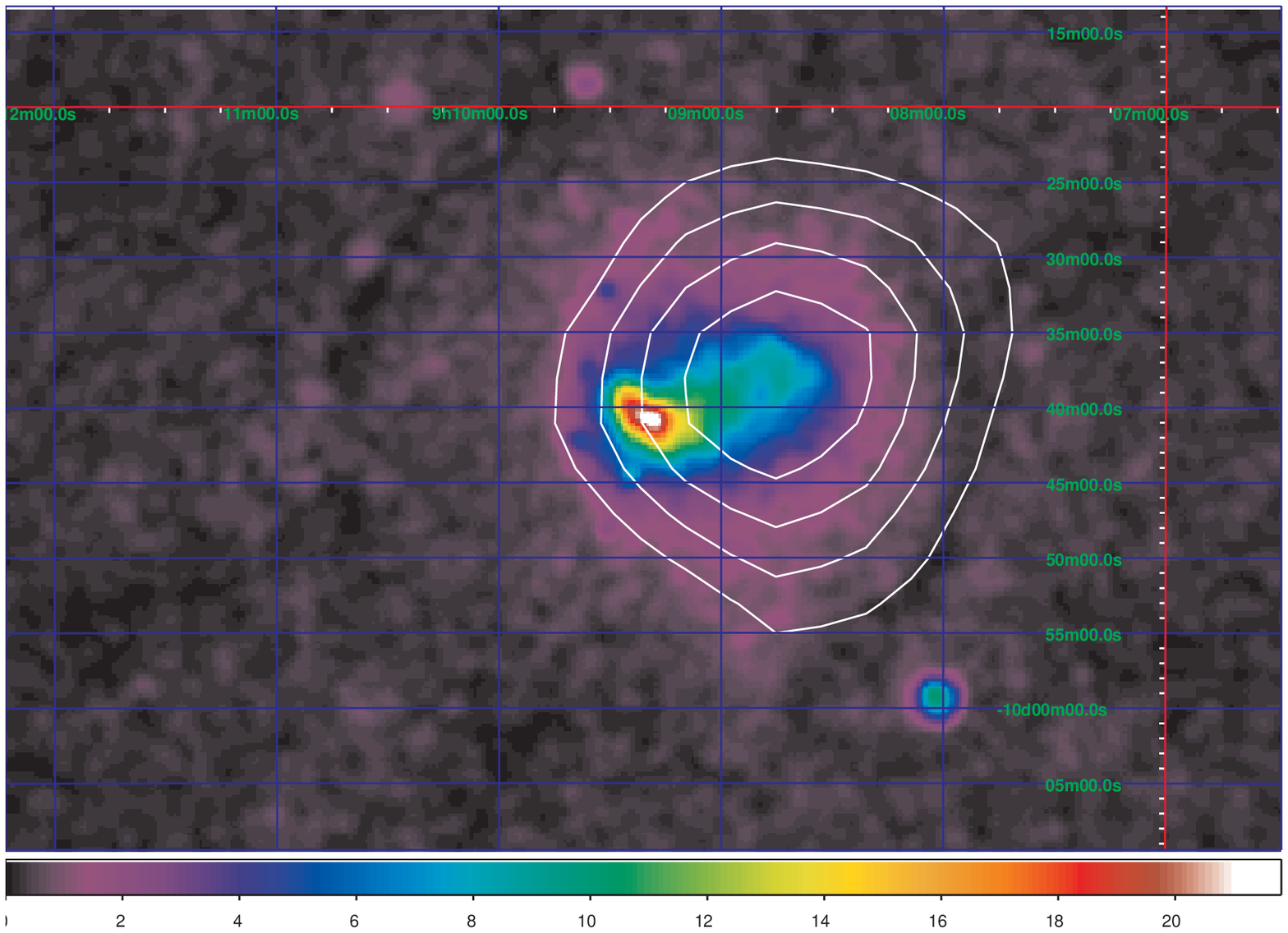} &
   \includegraphics[scale=0.45,angle=0]{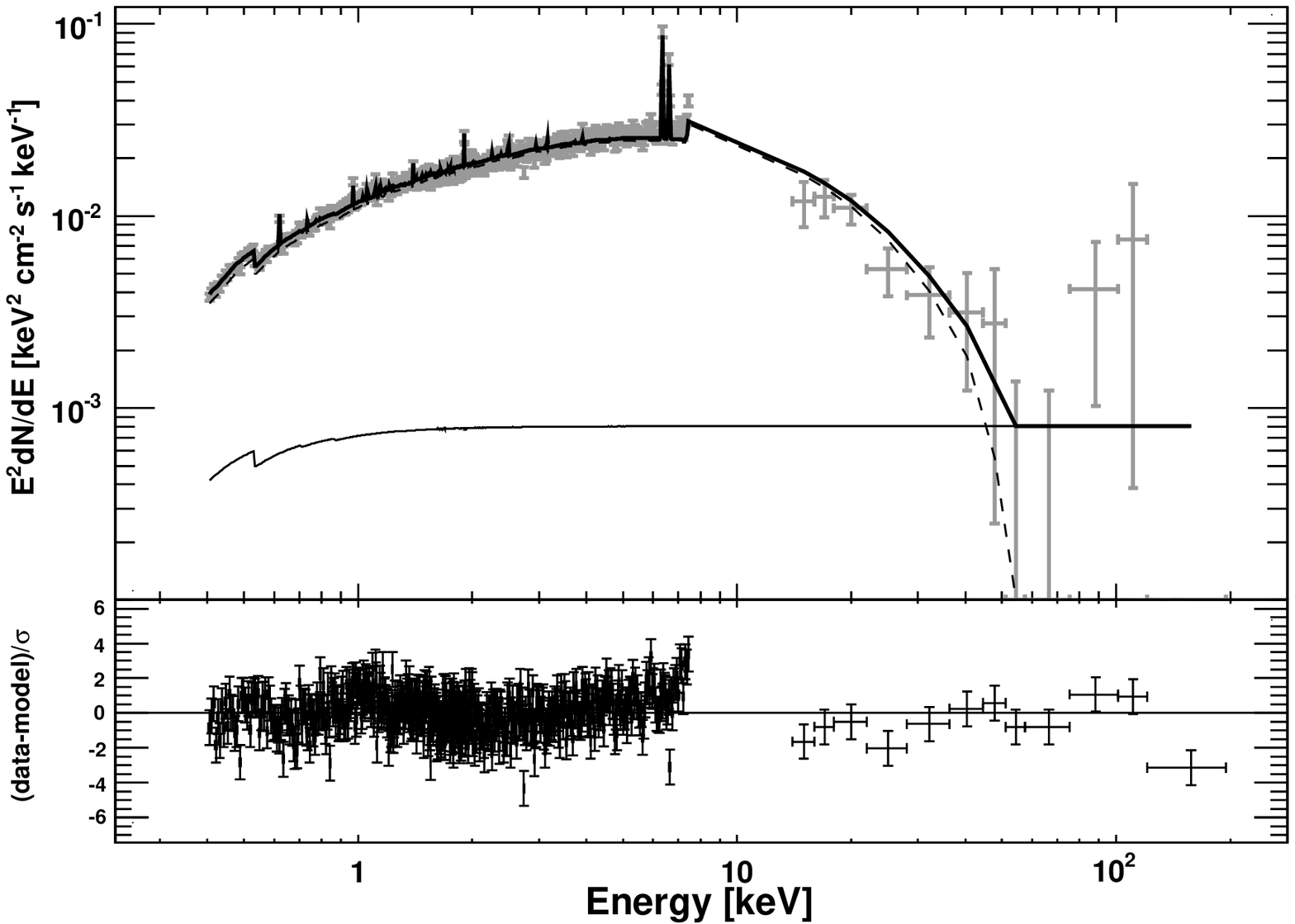} \\
    \end{tabular}
    \end{center}
    \null\vspace{-7mm}
   \caption{
{\it Left Panel:} ROSAT 0.1--2.4\,keV surface brightness of Abell 0754
with BAT significance contours superimposed. The contours range from 
2.5\,$\sigma$ to 8.0\,$\sigma$.
{\it Right Panel:}
Joint fit to XMM-Newton--BAT data. The best fit model 
(thick solid line) is the sum of a thermal model (dashed line) and of 
a power law (thin solid line). 
}
  \label{fig:abell0754}
\end{figure*}

{\bf Swift J0908.9-0938} is associated with the well studied cluster 
of galaxies
Abell 0754. X-ray maps indicate that A 0754 is far from hydrostatic 
equilibrium, 
experiencing a violent merger \citep[i.e.][]{hen95,hen96}.
Its detection by RXTE \citep{val99,revnivtsev04} and BeppoSAX \citep{fusco03} above 15\,keV   make the association of the cluster with the BAT source secure. 
While the RXTE detections do not measure any significant hard X-ray excess, 
BeppoSAX detects a hard tail with a significant deviation from the thermal component in the 50--70\,keV energy range.
It is worth noting that the BAT centroid\footnotemark{} 
falls $\sim$6\arcmin\ west of
the brightest  region of the cluster (see Fig.~\ref{fig:abell0754}).
\footnotetext{For an 8\,$\sigma$ detection, the expected
 maximum offset of the BAT
centroid is $\sim$2.5\arcmin\ \citep[see Fig.~10 in][]{ajello07a}.}
Chandra analysis of the gas temperature spatial distribution shows indeed that 
the BAT position corresponds  to regions of hot (T$\approx$10--15\,keV) gas 
\citep{markevitch03}. The analysis of XMM-Newton data confirms the existence
of hot regions in the west part of the cluster \citep{henry04}.
On the other hand, centroid shifts as a function of the waveband
are a common indication of a merging cluster \citep{oha04}.

 The BAT spectrum, shown in Fig.~\ref{fig:abell0754}, 
is well fitted ($\chi^2/dof = 6.3/9$) 
by a single bremsstrahlung model with a plasma temperature of  
9.9$^{+4.3}_{-2.6}$\,keV.
This is  in good agreement
with the temperature of 9.4$^{+0.16}_{-0.17}$\,keV reported by 
\cite{fusco03} and 9.0$\pm0.13$\,keV reported by \cite{val99}.
The {\it BeppoSAX} 10--40\,keV non-thermal flux of $\sim$1.6$\times 10^{-12}$\,erg cm$^{-2}$ s$^{-1}$ is consistent with the (90\,\%) 
upper limit from BAT of 
6.5$\times 10^{-12}$\,erg cm$^{-2}$ s$^{-1}$.

XMM-Newton observed A 0754 for 11\,ks in May, 2001.
The XMM-Newton--BAT data are well fitted by a single APEC model
with a plasma temperature of 8.5$^{+0.19}_{-0.13}$\,keV
and 0.29$\pm 0.03$ solar abundance. Adding a power-law model,
with photon index fixed to 2.0, improves the fit (F-test probability 
4.6$\times10^{-9}$). The best fit temperature is $9.3\pm0.4$\,keV and
the non-thermal 50--100\,keV flux is 
7.6$^{+2.4}_{-2.7}\times 10^{-13}$\,erg cm$^{-2}$ s$^{-1}$.
The non-thermal flux in the 10--40\,keV band is 
1.7$^{+0.2}_{-0.6}\times 10^{-12}$\,erg cm$^{-2}$ s$^{-1}$ and 
is in good agreement with the non-thermal flux measured by \cite{fusco03}.
However, \cite{fusco03} also discuss the possibility that the 
non-thermal flux be produced by the BL Lac object 26W20.
This object lies $\sim$24\arcmin\ away from the BAT centroid,
and outside the XMM-Newton field of view, thus we can rule out
that it is contributing to the detected non-thermal flux. 

However, we note that several point-like
objects appear in the XMM-Newton image and within 10\arcmin\ from the BAT
centroid. A simple hardness ratio analysis reveals that the
hardest object is located at R.A., Dec. = 09 09 13.7, -09 43 05.4.
The likely counterpart is 2MASS	09091372-0943047	 for which
beside the magnitude (bmag = 20.0) nothing else is known.
The XMM-Newton spectrum is extremely hard. It can be well represented, 
in the 0.1--10\,keV energy range,
 by an absorbed power-law with photon index of 1.23$^{+0.33}_{-0.24}$
and absorption of $5.6^{+5.4}_{-2.6}\times10^{21}$\,atoms cm$^{-2}$. 
The source flux
extrapolated to the 10--40\,keV band is (1.3$\pm0.3$)$\times10^{-12}$\,erg
cm$^{-2}$ s$^{-1}$. It is thus clear that this single source accounts
for the non-thermal flux detected both by our and \cite{fusco03} analyses.

\cite{val99} and \cite{fusco03} derive a lower limit for the magnetic field $B$ of $\sim 0.2~\mu$G and $\sim 0.1~\mu$G respectively. 
Our estimate of $B$, reported  in
Table~\ref{tab:ul},  uses the VLA observations from \cite{fusco03} 
($S_{1365~ \textrm{MHz}}= 86~ \textrm{mJy}$,  $\alpha=1.5$) 
and is consistent with the results of \cite{bacchi03} and of \cite{fusco03}.

\subsubsection{Coma}
\label{subsec:coma}
{\bf Swift J1259.4+2757} is associated with the Coma cluster which
is one of the best studied cluster of galaxies.
Coma (aka A 1656) is a particularly rich and symmetric merging cluster. 
It has been known as a diffuse X-ray and radio source since forty years 
\citep{fel66,for72,wil70}. The cluster hosts a powerful radio halo
\citep{feretti98} and 
both BeppoSAX \citep{fus99} and RXTE \citep[][]{rep01,rephaeli02}
revealed the existence of non-thermal hard X-ray emission.

However, the detection of this hard X-ray excess is still 
quite controversial.
Indeed, the positive BeppoSAX detections \citep{fus99,fusco04} 
of hard X-ray excess were challenged by \cite{rossetti04,rossetti07}.
According to \cite{rossetti07}, the significance of the
non-thermal excess changes (decreases) with the best-fit plasma temperature 
and only a certain set of assumptions (e.g. temperature of the ICM)
leads to a significant hard X-ray excess.
However, recently \cite{fus07}, using different 
software analyses and studying a large set of background observations,
were able to confirm their previous finding. 
Independently of the BeppoSAX results,
the RXTE   detection \citep[][]{rep01,rephaeli02} of the hard X-ray excess
remains unchallenged.

Lately, Coma has also been targeted by INTEGRAL \citep{eck07,lutovinov08}.
\cite{eck07} show, in their combined XMM-Newton--INTEGRAL analysis, the presence
of a hotter region (gas temperature of 12$\pm2$\,keV as compared to
7.9$\pm0.1$\,keV of the center) in the south-west region.
The authors favor the possibility that this emission is produced by IC 
scattering  because its spatial distribution overlaps the 
halo of radio synchrotron radiation.
\cite{lutovinov08} using INTEGRAL, ROSAT and RXTE data
showed that the global Coma spectrum is well approximated by a thermal
emission model only and 
found very marginal evidences (1.6\,$\sigma$) for hard X-ray excess.
Thus, in light of these results the evidences 
for non-thermal emissions in Coma seem not conclusive.

\begin{figure*} 
    \begin{center}
     \begin{tabular}{cc}
    \includegraphics[scale=0.4,angle=0]{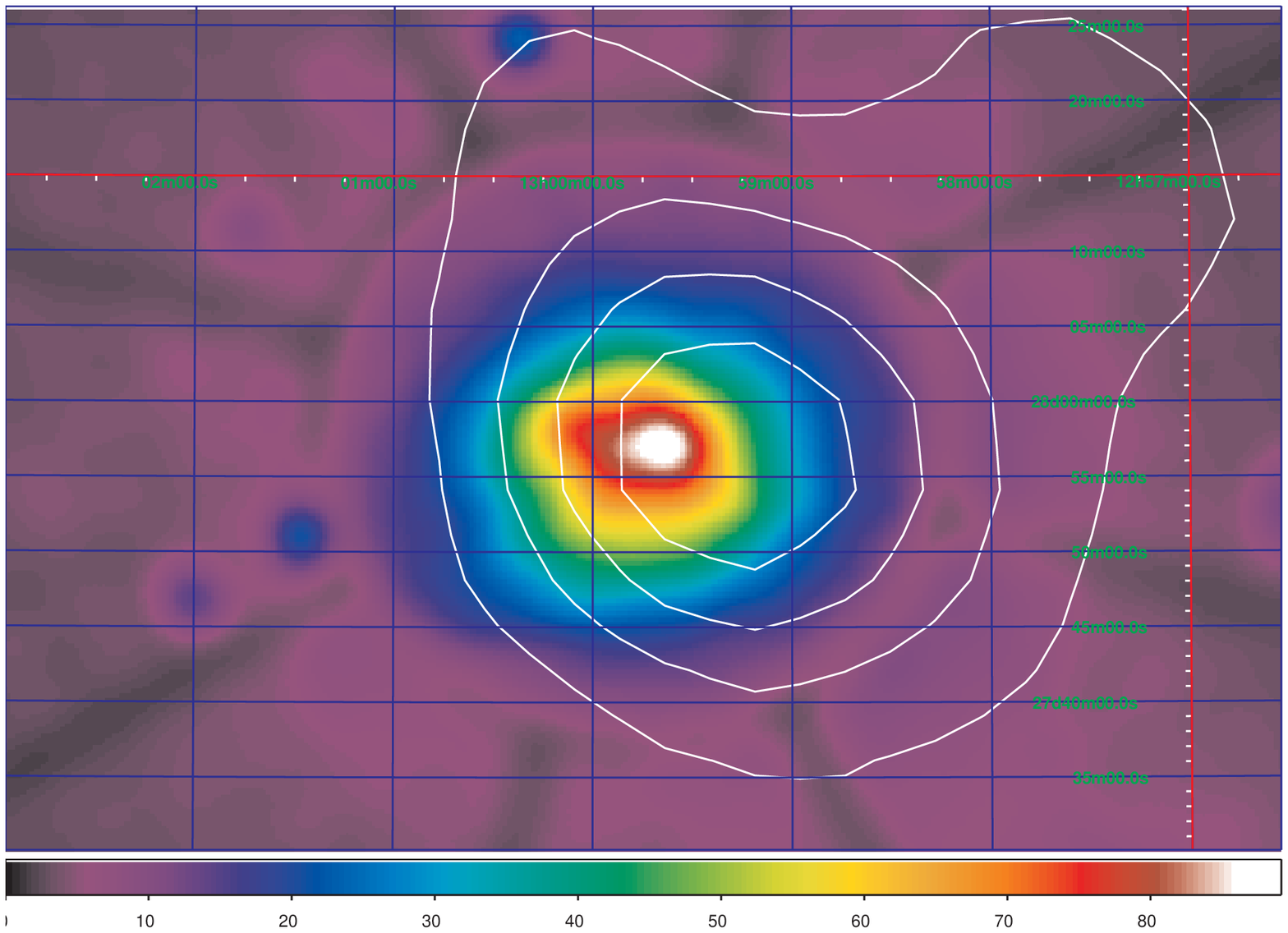} &
   \includegraphics[scale=0.45,angle=0]{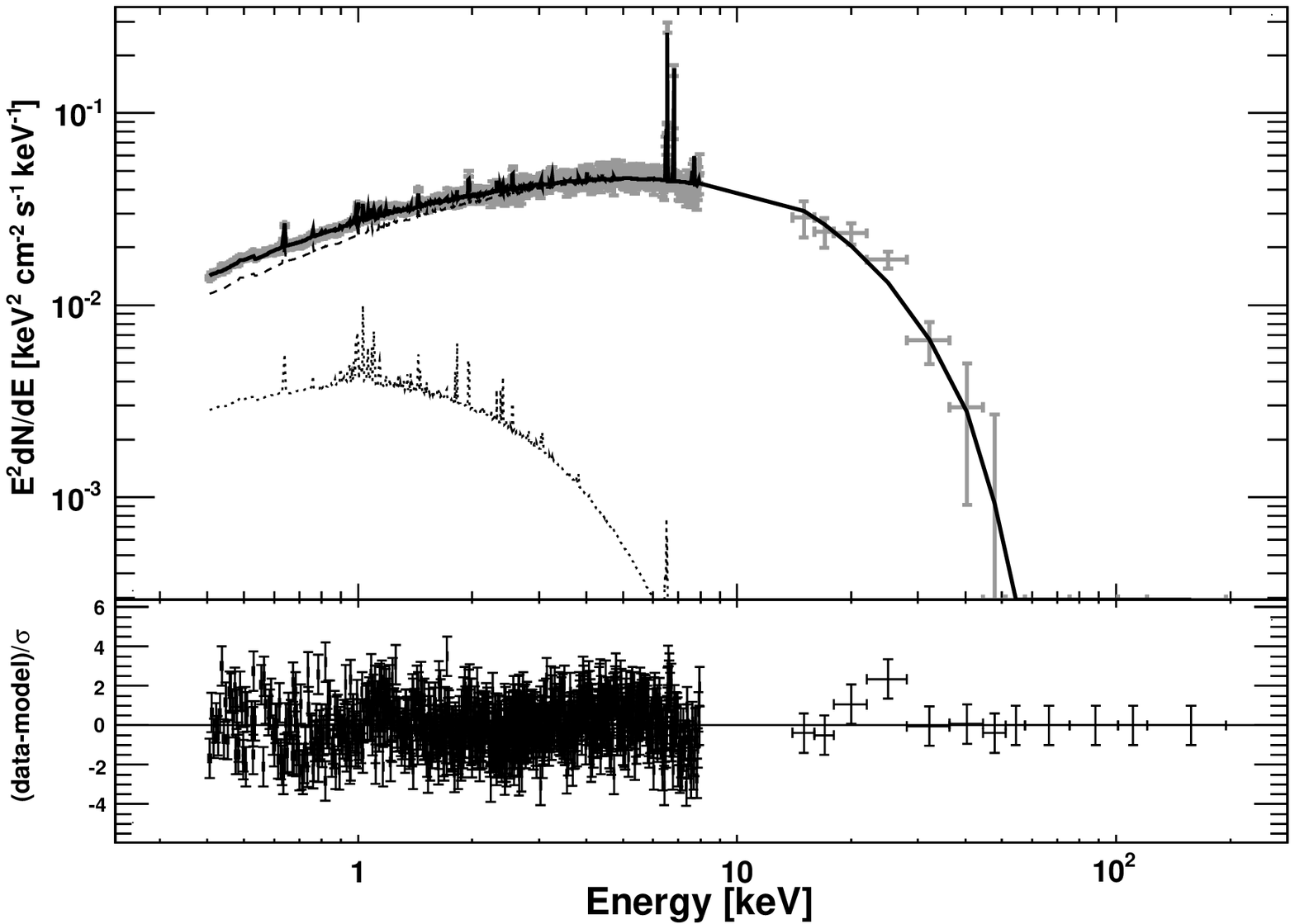} \\
    \end{tabular}
    \end{center}
    \null\vspace{-7mm}
   \caption{
{\it Left Panel:} ROSAT 0.1--2.4\,keV surface brightness of Coma
with BAT significance contours superimposed. The contours range from 
2.5\,$\sigma$ to 20\,$\sigma$.
{\it Right Panel:}
Joint fit to XMM-Newton--BAT data. The best fit model (solid line)
is the sum of two thermal models  (dashed and dotted lines).
}
  \label{fig:coma}
\end{figure*}

Coma is the only cluster in our sample whose extent is larger
than the BAT PSF. The analysis of point-like 
sources in the vicinity of the Coma cluster shows that the PSF full width
at half maximum (FWHM)
is 22\arcmin\ while the FWHM of the Coma detection 
is 26\arcmin\ . Using a simple Gaussian
profile for the surface brightness of Coma yields a 1\,$\sigma$ extent in the
10\arcmin--15\arcmin\ range. This is in agreement with the 
morphological analysis of \cite{eck07}. 
Moreover, from Figure~\ref{fig:coma}, the offset between
the BAT and the ROSAT centroids is apparent. Indeed, the BAT centroid falls $\sim$4\arcmin\
west of the ROSAT surface brightness peak. As discussed by
\cite{eck07} and \cite{lutovinov08} for INTEGRAL, the high-energy
centroid coincides with a region of hot gas likely due to an infalling 
subcluster.

Coded-mask detectors
suppress the flux of diffuse sources and in order to recover
the exact source flux and significance one needs to develop dedicated
methods for the analysis of extended objects \citep[e.g.][]{renaud06a}.
Given the fact that Coma is the only cluster 'resolved' by BAT,
a dedicated analysis will be left to a future paper \citep{ajello08d}.
However, we can extract the spectrum treating Coma as a point-like
source. This translates into an analysis of the source emission
within a radius of  $\sim$10\arcmin\ from the BAT centroid.
The BAT spectrum is well fitted by 
a thermal model with gas temperature of 9.13$^{+1.68}_{-1.31}$\,keV.

XMM-Newton observed Coma several times. We analyzed an observation
of 16\,ks which took place in June, 2005. The XMM-Newton spectrum
was extracted (as described in $\S$~\ref{subsec:individual}) 
including all photons within 10\arcmin\ from the BAT centroid.
Integrating the surface brightness profile derived by ROSAT
\citep[beta model with $\beta$=0.74 
and core radius R$_c$=10.7\arcmin ;][]{lutovinov08}
shows that 
our selection includes $\sim$75\,\% of the total Coma flux.
A fit to the XMM-Newton--BAT spectrum with a single-temperature 
model  does not yield satisfactory results ($\chi^2$/dof = 1168.9/858).
We then tried to add a power law to the APEC model.
Adding a power-law model improves the fit ($\chi^2$/dof = 905.5/856)
and results into a well constrained photon index of 2.11$^{+0.11}_{-0.13}$.
However, this fit leaves evident ('snake'-like) residuals at low energy
(see below for the residuals of all Coma fits).
These residuals might highlight the presence of another thermal
component. Indeed,
we find that a satisfactory fit ($\chi^2$/dof = 846.5/856) 
is achieved using two APEC models. The most intense component has a temperature
of 8.40$^{+0.25}_{-0.24}$\,keV and an abundance of 0.21$^{+0.03}_{-0.03}$
consistent with what found by \cite{arnaud01} and \cite{lutovinov08}.
The low-temperature component (T=1.45$^{+0.21}_{-0.11}$\,keV and 
Z = 0.05($\pm0.02$)Z$_{\odot}$)
accounts very likely for one or more of the X-ray sources in the field of Coma.
Indeed, an  hardness ratio analysis of these X-ray sources
shows that their spectra are compatible with thermal models with
temperatures in the 0.1--2\,keV range \citep{finoguenov04}.
According to \cite{finoguenov04}, these objects are 
(non-AGN) galaxies with a suppressed X-ray emission due to 
reduced star-formation activity.
Summarizing, we believe that the double-thermal model explains better the
data with respect the thermal plus power-law model because:
1) it produces the largest improvement in the fit (i.e. largest
 $\Delta \chi^2$),
2) it  better reproduces the low-energy part of the spectrum, and
3) it accounts for  all the point-like sources which are present in the 
XMM-Newton observation.
The best fit, sum of two APEC models, is shown in Fig.~\ref{fig:coma}.
The residuals to all the fits described in this section 
are reported in Fig.~\ref{fig:rescoma} while their parameters
are summarized in Tab.~\ref{tab:5}.

Our 99\,\% CL upper limit in the 50--100\,keV band is
1.70$\times 10^{-12}$\,erg cm$^{-2}$ s$^{-1}$.
However,
we remark that this spectrum is representative only of the 10\arcmin\ radius
region centered on the BAT centroid.
Indeed, since the IC and the thermal emissions are proportional 
to the electron density and to its square respectively 
\citep[F$_{IC} \propto n_e$ and F$_{thermal}\propto n^2_e$; e.g.][]{sarazin98},
their ratio (IC/thermal)  is expected to increase
with the distance from the cluster. Moreover, the lower density and 
larger sound speed (with respect the physical conditions in the core) 
make CR acceleration more efficient in the 
cluster outskirts \citep{pfrommer07}. 
For these reasons and because Coma is an extended source for BAT,
of which we analyze only the core, we cannot exclude the presence 
of  a non-thermal  component which arises in the outskirts of the cluster.

\subsubsection{Abell 3571}
{\bf Swift J1347.7-3253} is likely associated with the Abell 3571
cluster, which has also been detected in the RXTE Slew-Survey
\citep{revnivtsev04}. Its symmetric morphology, see left panel
of Fig.~\ref{fig:abell3571}, and temperature map
 indicate that A 3571  is a relaxed cluster \citep[e.g.][]{mar98}. 
However, the radio structure, of the complex in which A 3571 lays, suggests 
that this cluster is in the late stages of merging \citep{ven02}.
We note that  Abell 3571 is known to have a moderate
cool core \citep{peres98}.
Past and recent studies do not report evidences
for non-thermal hard X-ray emission in Abell 3571.
A fit to the BAT spectrum with a bremsstrahlung model yields a 
temperature of 6.9$^{+6.0}_{-2.6}$\,keV 
(in agreement with the mean temperature
of 6.71$^{+0.15}_{-0.42}$\,keV measured with Chandra by \cite{sanderson06}), 
but the chi-square ($\chi^2_{red}$=1.76) is relatively poor. 
The BAT spectrum shows positive residuals above 60\,keV which might reveal the presence
of a hard tail (see Fig.~\ref{fig:abell3571}). However,
given the low S/N of our spectrum, adding a power-law component
does not improve the chi-square.
XMM-Newton observed Abell 3571 for 12\,ks in July, 2007.
According \cite{nevalainen01}, the surface brightness of Abell 3571
follows a beta profile with $\beta$=0.68 and core radius R$_c$=3.85\arcmin\ .
Therefore, our region of 10\arcmin\ radius includes approximately
93\,\% of the cluster emission. This factor is taken into account when
performing the joint fit of XMM-Newton and BAT data.
The combined XMM-Newton--BAT spectrum, shown in the right panel 
of Fig.~\ref{fig:abell3571},
 is well fitted by an APEC model
with a plasma temperature of 6.01$\pm0.21$\,keV and an abundance
of 0.34$\pm 0.06$ solar.
The total 
2--10\,keV flux of (8.0$\pm$0.3)$ \times 10^{-11}$\,erg cm$^{-2}$ s$^{-1}$
is in good agreement with the value of 
7.3$\pm 0.4 \times 10^{-11}$\,erg cm$^{-2}$ s$^{-1}$ 
measured by BeppoSAX \citep{nevalainen01}. 
Even though statistically not required, a non-thermal power-law
(photon index fixed to 2.0) is well constrained by our data. Indeed,
we are able to derive a 50--100\,keV flux of 
(1.4$\pm0.5$)$\times10^{-12}$\,erg cm$^{-2}$ s$^{-1}$.

The radio flux density from the NRAO VLA Sky Survey is 
$S_{1380 ~\textrm{MHz}}= 8.4 ~\textrm{mJy}$
 \citep{con98}. We could not find any reference for the spectral 
index, so we adopted the value of $\alpha=1.5$ which leads to 
the lower limit listed in Table~\ref{tab:ul}.
We note that steeper spectrum gives a larger upper 
limit for the magnetic field (e.g. $\alpha=2$) would
yield a lower limit twice as large as the previous.

\subsubsection{Abell 2029}
{\bf Swift J1511.0+0544} is likely associated with the Abell 2029
cluster, which has also been detected at high energy by RXTE,
 BeppoSAX and Chandra
\citep[][respectively]{revnivtsev04,molendi99,clarke04}.
Left panel of Figure~\ref{fig:abell2029} shows that
the BAT source is well centered on the cluster emission as 
seen by ROSAT.
Abell 2029 has a moderate cool core \citep{sar98,molendi99}. 
\cite{clarke04} present an analysis of Chandra observations of 
the central region  and find signs of interactions between 
the X-ray and the radio plasma. The unusual central radio source 
(PKS0745-191) morphology would be typical of a merging cluster. 
They suggest that A2029 is a cluster that started very recently 
to cool to lower temperatures.

The BAT data alone are well fit ($\chi^2/dof=6.89/10$) 
by a simple bremsstrahlung
model with a temperature of 10.6$^{+5.8}_{-3.3}$\,keV.
An 8\,ks long XRT observation took place in September, 2005.
Given the extent of the XRT CCD, we extracted all the photons within
a 6\arcmin\ from the BAT centroid. The surface brightness
profile follows a beta model 
with $\beta$=0.64 and core radius R$_c$=1.8\arcmin\  \citep{sarazin98}.
Integrating the beta profile up to 6\arcmin\ yields that 95\,\%
of the total cluster emission is included in our selection.
However, for the case of Abell 2029 the beta profile fails
to explain the inner 1.8\arcmin\ region which is characterized by a
bright core \citep{sarazin98}. Thus, our selection might include an higher
fraction of the total cluster emission. Indeed, BAT and XRT data are
well fitted without the needs of a cross-normalization constant.
The BAT and XRT data are successfully fitted,
 by an APEC model with 
plasma temperature of 7.45$\pm0.34$\,keV and 0.39$\pm0.09$ solar abundance
which is consistent with the Chandra results \citep{clarke04}.
From the combined fit, we derive a 99\,\% CL upper limit
to the non-thermal flux in the 50--100\,keV band of 1.27$\times10^{-12}$
erg cm$^{-2}$ s$^{-1}$.
However, we note that the fit leaves positive residuals at high energy.
We thus used a second APEC model, with abundance fixed
at 0.4,  to account for them.
The F-test confirms that the second thermal component is detected at 
99.85\,\% CL.  The best fit temperatures are 9.6$^{+2.0}_{-2.0}$\,keV
and 4.1$^{+1.7}_{-1.5}$\,keV respectively.
This fit is shown in Fig.~\ref{fig:abell2029}.
Abell 2029 has been targeted by ground-based TeV telescopes; however
no TeV emission has been detected so far \citep{perkins06}.

\cite{con98} found $S_{1380 ~\textrm{MHz}}= 527.8 ~\textrm{mJy}$. 
We adopted the value of $\alpha=1.5$ which leads to
the lower limits on the magnetic field estimated in Table~\ref{tab:ul}.
We note that \cite{tay94} obtained a lower limit on the magnetic field
of 0.18$\mu$G using observations of the central radio galaxy.

\subsubsection{Abell 2142}
{\bf Swift J1558.5+2714} is associated with the Abell 2142 merging cluster.
The detection in the 3--20\,keV band  by RXTE \citep{revnivtsev04}
makes the association of the BAT source with the cluster rather
strong. According to \cite{peres98} and \cite{mol02}, Abell 2142 has a 
cool core that survived the merger.
 \cite{markevitch00} and \cite{sanderson06}, using Chandra observations, 
noted that the core of A2142 has  a complex structure, probably with 
a poor cluster enclosed in the halo of a hotter larger cluster. This 
would explain the lower temperature in the center, without the presence 
of a cool core. 
The left panel of Fig.~\ref{fig:abell2142} shows a point-like source
located $<$4\arcmin\ from the cluster center. This object is the 
Seyfert 1 galaxy 2E 1556.4+2725. Given the distance, both objects,
the cluster and the Sy1, are not separated by BAT.

The BAT data are well fit by a simple bremsstrahlung model 
($\chi^2/dof=7.96/10$) with plasma temperature of 10.1$^{+3.7}_{-2.7}$\,keV.
We analyzed an XMM-Newton observation of 800\,s in conjunction with the BAT data.
In this case, we extracted separately the spectrum of the cluster
and the spectrum of the Sy1  2E 1556.4+2725.
The latter one shows an X-ray spectrum typical of a Sy1 object: 
i.e. absorption consistent with 
the Galactic one and photon index of 1.98$^{+0.16}_{-0.14}$.
The extrapolated flux in the 15-55\,keV range is 
2.3$\times10^{-12}$erg cm$^{-2}$ s$^{1}$ and it is well below 
the BAT sensitivity. Therefore we can consider negligible 
the Sy1 contribution in the BAT band.
The surface brightness of Abell 2142
profile follows a beta model 
with $\beta$=0.83 and core radius R$_c$=4.2\arcmin\  \citep{henry96}.
Integrating the beta profile up to 10\arcmin\ yields that 97\,\%
of the total cluster emission is included in our selection.
However, for the case of Abell 2142 the beta profile underestimate
the brightness of  the inner 3\arcmin\ region which is characterized by a
bright core \citep{henry96}. Thus, our selection might include an higher
fraction of the total cluster emission. Indeed, BAT and XMM data are
well fitted without the need of a cross-normalization constant.
The cluster XMM-Newton--BAT spectrum is well fit by a simple APEC model with a plasma
temperature of 8.40$^{+0.64}_{-0.45}$\,keV. The fit is shown in the right panel
of Fig.~\ref{fig:abell2142}.
This is
well in agreement with  the temperatures of 8.8$^{+1.2}_{-0.9}$\,keV 
and 9.0$\pm0.3$\,keV  measured by Chandra and GINGA respectively
\citep{markevitch00,white94}. From our fit the abundance is 
 0.27$^{+0.13}_{-0.13}$ solar.
Since no hard X-ray excess is detected, we report
99\,\% CL upper limits. 
Using a power-law with photon index of 2.0, we derive from
the XMM-Newton--BAT data a 99\,\% CL upper limit to the 50--100\,keV
non-thermal flux of 1.6$\times10^{-12}$\,erg cm$^{-2}$ s$^{-1}$.
The 99\,\% CL limit on  the non-thermal luminosity is 
6.1$\times 10^{43}$\,erg s$^{-1}$.
The marginal ($\sim2$\,$\sigma$) {\it BeppoSAX} detection 
of a non-thermal emission  \citep{nevalainen04} is a factor
5 larger than our upper limit and thus incompatible with our
data.

The presence of a radio halo was already reported by \cite{har77}. \cite{gio00} measured
 $S_{1400~ \textrm{MHz}}= 18.3~ \textrm{mJy}$. 
In absence of a measured index $\alpha$ we adopt 
the arbitrary value of $\alpha=1.5$ to obtain the magnetic field
constraint listed in Table~\ref{tab:ul}.

\subsubsection{Triangulum Australis}
{\bf Swift J1638.8-6424}, shown in the right panel of 
Fig.~\ref{fig:triangulum},
 is likely associated with the hot X-ray cluster of galaxies
Triangulum Australis.
This cluster at z=0.058 has already been detected in the
 ROSAT, RXTE Slew  and INTEGRAL surveys 
\citep{voges99,revnivtsev04,stephen06}. 
In particular the detections by RXTE and INTEGRAL above 15\,keV make this 
association certain.
The Triangulum Australis cluster may host a cool core \citep{edg92,peres98}.
However, \cite{mar96} used the temperature and entropy maps from ASCA 
and ROSAT to find an indication of the probable presence of a subcluster
 merger and argue that the cool gas in the core does not require a cooling
 flow. \cite{mar98} found that a non-thermal component is more likely than 
a cooling flow.

The BAT spectrum is well fitted ($\chi^2/dof=5.68/9$) by a simple
bremsstrahlung model with plasma temperature of 13.4$^{+6.3}_{-3.7}$\,keV.
A similar temperature was found by \cite{mar96} in the centre of the cluster.

XMM-Newton observed the Triangulum Australis cluster for 7480\,s 
in February, 2001. According to the beta profile reported by \cite{mar96},
selecting photons within 10\arcmin\ of the BAT centroid includes 
$\sim$92\,\% of the cluster emission. We thus employ such cross-normalization
factor when fitting XMM-Newton and BAT data.
The BAT and XMM-Newton data are consistent with a pure APEC model.
From the best fit, shown in the right panel of Fig.~\ref{fig:triangulum},
we derive a plasma temperature of 9.30$^{+0.30}_{-0.30}$\,keV
and an abundance of 0.30$^{+0.07}_{-0.07}$ solar.
The XMM-Newton--BAT temperature is in agreement 
with the mean values of 9.06$^{+0.33}_{-0.31}$\,keV and   9.50$\pm0.70$\,keV
reported by \cite{ikebe02} and by \cite{chen07} respectively.
Using a power-law with photon index of 2.0, 
we derive a 99\,\% CL upper limit to non-thermal
emission in the 50--100\,keV band of 
6.5$\times 10^{-13}$\, erg cm$^{-2}$ s$^{-1}$.

\cite{condon93} report a  4.85\,GHz radio source centered $\sim$7\arcmin\
 away from the  BAT centroid. They find an upper
limit of $33$\,Jy. We adopt this flux and
the arbitrary value of $\alpha=1.5$ to obtain the magnetic field 
constraint listed in Table~\ref{tab:ul}.

\subsubsection{Ophiucus}
{\bf Swift J1712.3-2319}  lays only 1.7\arcmin\ (see Fig.~\ref{fig:ophiucus})
 away from one of the most
studied galaxy clusters, Ophiucus, discovered by \cite{joh81}. 
The detection at high-energies by BeppoSAX and INTEGRAL 
\citep[][respectively]{nevalainen01,bird06}
makes the association with the BAT source  certain.
\cite{wat01} used ASCA to measure the X-ray brightness distribution 
and temperature map. Considering the similarities with the Coma cluster, 
they conclude that Ophiucus is not relaxed and  has likely experienced a 
recent merger.  
The BAT-derived plasma temperature of 9.5$^{+1.4}_{-1.1}$\,keV
is in good agreement with the values of
 9.6$^{+0.6}_{-0.5}$\,keV and 9.0$^{+0.3}_{-0.3}$\,keV 
measured by BeppoSAX \citep{nevalainen01} and by Suzaku \citep{fujita08}.\\
An hard X-ray excess was detected by \cite{nevalainen01} at a 
2\,$\sigma$ level.
Very recently, \cite{eck07b} using INTEGRAL confirmed this   hard X-ray emission
at an higher confidence level (4--6.4\,$\sigma$). 
The imaging capabilities of the instruments on-board INTEGRAL 
allowed the authors to conclude that the observed excess 
over the thermal emission
is not originating from point sources (such as obscured AGNs) and is 
therefore non-thermal. 
This excess is marginally consistent with BAT data.
Indeed, from our data we derive a 90\,\%  upper limit to the non-thermal
component (20--60\,keV) of  7.2$\times 10^{-12}$\,erg cm$^{-2}$
s$^{-1}$ while the reported non-thermal flux 
observed by INTEGRAL is 
(10.1$\pm2.5$)$ \times 10^{-12}$\,erg cm$^{-2}$ s$^{-1}$.

We analyzed an archival Chandra observation of $\sim$50\,ks. The 
observation, which took place in October 2002, was performed using the  ACIS-S.
Given its extent, the Ophiucus cluster is not entirely contained in a
single chip. We thus extracted only those photons in a region
of radius of  2.1\arcmin\ around the BAT centroid. The region extent
is dictated by the size of the chip. When performing a simultaneous fit
with BAT data, we must therefore account for the flux which falls outside
of the ACIS-S chip. Assuming that the surface density follows a beta
profile and adopting the values of $\beta=$0.64 and core radius of 
R$_c$=3.2\arcmin\ as found by \cite{watanabe01} and confirmed
by \cite{eck07b}, we derive that only 
$\sim$52\,\% of the total cluster flux is included in our selection.
If we let the cross-normalization of the BAT and the Chandra data
vary, we derive that the Chandra data show a normalization (with respect
the BAT ones) of 53$^{+5}_{-6}$\,\% which is in good agreement with the 52\,\%
derived above. Thus, we fix the cross-normalization factor
at 52\,\%.  Moreover, as in \cite{blanton03} we accounted for 
the uncertainty in the background subtraction adding
a systematic uncertainty of 2\,\%.
The joint Chandra--BAT spectrum is well fitted by a single
APEC model with a temperature of 9.93$^{+0.24}_{-0.24}$\,keV and 
abundance of 0.52$\pm0.03$.
Using a power-law with photon index of 2.0, 
we derive a 99\,\% CL upper limit to the non-thermal
emission in the 50--100\,keV  and 20--60\,keV bands of 
2.8$\times 10^{-12}$\, erg cm$^{-2}$ s$^{-1}$ and 
4.5$\times 10^{-12}$\, erg cm$^{-2}$ s$^{-1}$ respectively.
The INTEGRAL detection is inconsistent ($\sim$2\,$\sigma$) with our
upper limit.

The Ophiucus cluster is associated in the radio domain to 
the extended radio source MSH 17-203 \citep{joh81}. The most 
recent radio data date back to 1977 \citep{sle77} and report 
$S_{160 ~\textrm{MHz}}= 6.4$\,Jy and $\alpha=2$, which
we use to produce the lower limit on the magnetic field reported
in Tab.~\ref{tab:ul}. The results do not change if we use older
radio measurements \citep[e.g.][]{mil60,jon74,sle75}.

\subsubsection{Abell 2319}
{\bf Swift J1920.9+4357} is certainly associated with the 
massive Abell 2319 cluster,
 that is undergoing a major merger \citep[e.g][]{oha04}.
The BAT centroid (see left panel of Fig.~\ref{fig:abell2319}) lies
$\sim$2\arcmin\ north-west of the peak of the ROSAT emission.
Indeed, Chandra observations reveal at the same position a region
of hot, $\sim$12\,keV, gas while at the position of the ROSAT
peak there is likely a cool core \citep[][]{oha04}.
Abell 2319 has been detected above 10\,keV by BeppoSAX and RXTE 
\citep[][respectively]{mol99b,gruber02}. These two measurements
are symptomatic of the uncertainty related to the hard X-ray 
detection claims from non-imaging instruments and the inherent
uncertainty from source contamination.
Indeed, \cite{molendi99} report that no hard-tail emission
is present in BeppoSAX data, while \cite{gruber02} find that a power-law
component can explain some residual features in the 15--30\,keV energy range.
The BAT data favors the thermal scenario. Indeed the best fit to the data
is obtained using a  pure bremsstrahlung model with 
a plasma temperature of 14.1$^{+4.0}_{-3.0}$\,keV consistent,
within the large errors,  with 
the 9.6$\pm0.3$\,keV  value measured by BeppoSAX.

In addition, we analyzed a 10\,ks XMM-Newton observation together 
with the BAT data. Utilizing the surface brightness profile
obtained by \cite{oha04} (beta model with $\beta$=0.55 and core radius
R$_c$=2.6\arcmin\ ) we determine that our region of 10\arcmin\ radius  
includes $\sim$90\,\% of the cluster emission. We employ such
cross-normalization factor when fitting XMM-Newton and BAT data.
The BAT--XMM-Newton spectra, 
shown in the right panel of Fig.~\ref{fig:abell2319},
 are well fitted by an APEC model with
a plasma temperature of 9.27$^{+0.27}_{-0.27}$\,keV and abundance
of 0.25($\pm 0.04$) solar. The 99\,\% upper limit on the 
2--10\,keV non-thermal flux of 2.70$\times 10^{-12}$ erg cm$^{-2}$ s$^{-1}$ 
is in disagreement with  the non-thermal flux of 
(4.0$\pm0.1$)$\times 10^{-11}$\,erg cm$^{-2}$ s$^{-1}$ detected
in the same band by RXTE \citep{gruber02}.

\cite{har78} discovered a diffuse radio halo associated with the 
A 2319 cluster. 
An intensive study was done by \cite{fer97}, from which we take
 $S_{610 ~\textrm{MHz}}= ~1\, $Jy and $\alpha=0.92$
to estimate the lower limit on the magnetic field reported in 
Table~\ref{tab:ul}.

%
%
\section{Clusters Properties}
\subsection{Constraints on non-thermal excess emission}
\label{subsec:non-thermal}
In order to constrain the non-thermal hard X-ray emission,
we have produced 3\,$\sigma$ upper limits on the 50--100\,keV non-thermal
flux for each source presented in the previous section.
We excluded the Perseus and the Coma clusters. Indeed, Perseus
is the only cluster where the detected ``hard-tail'' is certainly
produced by the brightest AGN while Coma requires a dedicated analysis.
We chose the 50--100\,keV energy band because above 50\,keV the thermal
emission of the clusters is negligible.

The 3\,$\sigma$ upper limit has been computed integrating the source
flux in the 50--100\,keV range and subtracting the thermal flux arising
from the best thermal fit. We added to this value three times the 1\,$\sigma$
uncertainty. The upper limits are reported in Table \ref{tab:ul}.
These upper limits were derived using BAT data alone.
It is important to note that, indeed, thanks to the very good sensitivity
of BAT, all these upper limits are very stringent. Indeed, the non-thermal
flux for all these sources is constrained to be below $\sim$1\,mCrab.

In the derivation above, we do  not make any assumption on the mechanism
generating the non-thermal flux. However, in most cases IC scattering
is believed to be the principal emission process 
\citep[e.g.][]{sar99,nevalainen04,fus07,eck07b}.
If this is true, then the IC emission can be modeled as 
a power-law with photon index $\sim$2 in the 1--200\,keV energy 
range \citep[see e.g.][]{reimer04}.
We thus computed the 99\,\% CL upper limits to the IC flux in the 50--100\,keV
band by 
adding a power-law model to the best fits reported in Tab.~\ref{tab:spec}. 
These limits are reported in Tab.~\ref{tab:ulxmm}.
It is worth noting that, since we are using XMM-Newton/XRT/Chandra
 and BAT data,
these  upper limits are a factor 5--10 lower than
those derived using BAT data alone (see Tab.~\ref{tab:ul}).

%
%

\subsection{Stacking analysis}
\label{subsec:stack}

A few clusters show positive, marginal, residuals above 50 keV; this is 
the case for Abell 3266, Abell 3571 and Abell 2142. 
Such features are not statistically significant to warrant an 
additional component (e.g. non-thermal power-law).
However, it might be that the non-thermal component be just
below the BAT sensitivity for such clusters.
In this case the stacking technique offers the capability to explore
the average properties of a given population beyond the current instrumental
limit. Thus, we produced the stacked spectrum of all clusters
except Perseus and Coma (for the reasons explained above). 
The average spectrum is produced by the weighted average
of all the spectra. The weight is chosen to be the inverse of the variance
of a given bin and it is exactly the same procedure used to extract
the spectra of each individual source. The same stacking
technique has been applied with success to the study of Seyfert galaxies
detected by BAT \citep{ajello07b}.
The total spectrum has an exposure time of $\sim$56\,Ms and it is shown
in Fig.~\ref{fig:stacked}. 
A fit with a simple bremsstrahlung model yields a
good chi-square ($\chi^2$/dof = 7.2/10).
The best fit temperature is 10.8$^{+0.9}_{-0.8}$\,keV  which is in
very good agreement with the mean temperature of 10.4\,keV as 
determined by averaging the values obtained fitting a simple
bremsstrahlung model to each cluster's spectrum (using BAT data alone).
This is a good confirmation that the chosen stacking technique
reproduces well the average properties of our cluster sample.

From the best thermal fit, we derive a 99\,\% CL upper limit (50--100\,keV)
for the non-thermal component of 1.9$\times 10^{-12}$\,erg cm$^{-2}$ s$^{-1}$
(0.3\,mCrab). At the average redshift of the sample (z=0.058),
this translates into a limiting luminosity of 
1.4$\times 10^{43}$\,erg s$^{-1}$.
\cite{nevalainen04} reported the detection of an average non-thermal 
component detected in the stacked spectrum (20--80\,keV) of {\it BeppoSAX} 
clusters. Their non-thermal luminosity is comprised 
\footnotemark{}
in the  (0.5--5.0)$\times 10^{43}$\,erg s$^{-1}$ range.
\footnotetext{The measurement
reported by \cite{nevalainen04} had to be converted to the
Hubble constant used in this paper.}
In the 20--80\,keV band our 99\,\% CL limit on the non-thermal luminosity
is 2.2$\times 10^{43}$\,erg s$^{-1}$. 
Thus, the findings of \cite{nevalainen04} are consistent
with our analysis.

On the other hand, all clusters, except perhaps Perseus and Abell 3571,
are undergoing a  merging phase. These last two
clusters are those which show the lowest ICM temperature in our sample.
The $L_x-T $ relation (shown in Fig.~\ref{fig:lx_vs_T})
reinforces the picture that most of the BAT clusters are mergers.
Indeed, the best fit to the data with a power-law of the form 
$L=A_6 T^{\alpha}_6$, where $T_6=T/6$\,keV
(fixing $\alpha$ at 2.88\footnotemark{}) yields a normalization 
$A_6$=(2.82$\pm$0.8)$\times 10^{43}\ h_{70}^{-2}$\,erg s$^{-1}$.
\footnotetext{Given the small range in luminosity spanned by our
sample we fixed $\alpha$ at the value determined by \cite{arnaud99}.}
While, \cite{markevitch98}  and  \cite{arnaud99},
found  for  $A_6$ a value of 
(12.53$\pm1.08$)$\times 10^{43}\ h_{70}^{-2}$\,erg s$^{-1}$  and 
(12.13$\pm0.06$)$\times 10^{43}\ h_{70}^{-2}$\,erg s$^{-1}$ respectively.
Indeed, merging clusters are known to segregate at lower luminosities
(or higher temperatures) in the $L_x-T$ plane \citep{ota06}.

There is a growing evidence which points towards a rather
non-uniform distribution of temperatures in the ICM of merging clusters
\citep[e.g.][]{markevitch03,oha04,eck07}. Both hydrodynamical
simulations \citep[e.g.][]{takizawa99} and observations 
\citep[see][for Abell 0754]{markevitch03} have shown that 
shocks due to cluster mergers can heat the ICM up to $\sim$15\,keV. 
Figure~\ref{fig:temp} shows that for the 
merging clusters the mean temperature measured by BAT is slightly
larger (given the large uncertainties)
than the mean ICM temperature measured below 10\,keV.
A similar trend, although using different wavebands,
has been recently reported  for  a sample 
of 192 galaxy clusters \citep{cavagnolo08}.
Moreover, for the merging clusters, the BAT centroid is shifted to positions
where Chandra and XMM-Newton have detected regions of hot gas.
Based on these evidences, we 
believe that the conjecture that these clusters show
regions of ``hot'' gas is a more viable claim than the one which
foresees the  presence of a strong IC component.

This claim is also supported by the fact that the high-energy
residuals (e.g. residuals above 10\,keV of the spectral fits
using a single thermal model) are in general better described
by an additional thermal component than a power-law model.
To prove this, we selected those clusters which show, in the
analysis presented in $\S$~\ref{subsec:individual}, the largest
residuals above 10\,keV from the thermal model used.
These clusters, which are Abell 2029, Triangulum Australis and Abell 2319,
also show a large deviation between the ICM temperature measured
below and above 10\,keV (see Fig.~\ref{fig:temp}).
We made a fit to  each of these clusters with: 1) a single thermal model,
2) the sum of a thermal and a power-law model,
and 3) the sum of two thermal models. 
The residuals to each of these fits are shown in Figures~\ref{fig:res2}, \ref{fig:res3}, and \ref{fig:res4}
 while the spectral parameters are 
summarized in Table~\ref{tab:5}. We note that in all three cases the
additional thermal model explains the residuals better than an additional
power-law model. We also remark that for most of the BAT clusters 
(in this case for Triangulum Australis and Abell 2319) the single thermal
model is already a good description of the data ($\chi^2$/dof$=\sim$1.0)
and given the statistics no other additional model is required.
This means that currently the high-energy residuals (with respect 
a single thermal fit) are not significant. Longer BAT exposures 
will clarify  the existence and nature  of these emissions.

%
%

\subsection{Cluster Magnetic Field Assessment}
\label{subsec:magnetic}

The diffuse synchrotron radio emission (radio halos, relics and mini-halos) proves the existence of magnetic fields in the ICM. The intensity of the synchrotron emission depends both on the strength of the magnetic field and on the electron density.
If the non-thermal X-ray emission results from IC  scattering off the same radio electrons by the CMB, then the degeneracy in magnetic field and relativistic electron density can be broken  \citep[e.g.][]{rep01}. 
Therefore the non-detection of a non-thermal component can be used to place a lower limit on the magnetic fields $B$ in clusters (the ratio of IC to radio flux is inversely proportional to $B^{\alpha+1}$). 
Following \cite{har74} and \cite{sar88},
we estimate the lower limit on $B$ (the volume averaged component along the line of sight):

\begin{equation}
\frac{f_x ~\nu_r^{-\alpha }}{s_r~ \left(\int_{\nu_{min}}^{\nu_{max}} \nu_x^{-\alpha} d\nu_x \right)} = \frac{2.47 \times 10^{-19}~  T_{CMB}^3 b(p) }{B~ a(p)}\left(\frac{4960~ T_{CMB}}{B}\right)^{\alpha },
\end{equation}

where $\alpha$ is the spectral index, $p=2~\alpha+1$, $f_x$ the  
X-ray flux integrated over the band between $\nu_{min}=50$ KeV and 
$\nu_{max}=100$ KeV 
($f_x=k_c~\int_{\nu_{min}}^{\nu_{max}} \nu_x^{-\alpha} d\nu_x$, 
in erg cm$^{-2}$ s$^{-1}$), $s_r=k_s~\nu_r^{-\alpha} $ the flux density 
at the radio frequency  $\nu_r$ (in erg cm$^{-2}$ s$^{-1}$ Hz$^{-1}$),
 $T_{CMB}=2.7 K$ the temperature of the CMB  and $a(p)$ and $b(p)$ 
as in \cite{sar88} (eq. 5.6 and 5.8).
Since our clusters are nearby, in the formula above
we neglect redshift corrections.

Although the limit on the X-ray flux is very stringent, the measurement of 
the diffuse radio emission is complicated by the presence of individual
 radio galaxies in the cluster. 
In most cases the radio observations were not sensitive enough over a wide 
range of spatial scales to subtract the contribution of the single sources. 
Moreover,  the spectral index varies with the distance from the center 
of the cluster.
These factors make the  derivation of the magnetic field intensity uncertain.
 Therefore the values listed in table 3 have to be taken as order of magnitude
 estimates. Such estimates point to magnetic fields that are typically a
 fraction of a $\mu$G. These  low values
indicate that these systems are far from equipartition. This is possible if
 one considers that the magnetic fields and the relativistic particles may 
have a different spatial extension and history.

The magnetic field can also be evaluated by measuring the Faraday rotation 
(RM) of the plane of polarization from the radio galaxies in the cluster or
 in the background \citep[e.g.][]{kim91,clarke01}. 
The two estimates are different 
(with $B_{RM}\gg B$), most likely because the interpretation of Faraday 
rotation measurements and the derivation
of the mean magnetic field strength rests on assumptions of the magnetic
field topology \citep[see][for an extensive discussion]{gol93,col05}.
We can produce a more robust, upper limit on the IC flux considering
that the IC emission spectrum can be approximated as a power law
in the 1--200\,keV energy band \citep[see e.g.][for more details]{reimer04}.
Using both 2--10\,keV and BAT data we were able to produce the limits
reported in Table~\ref{tab:ulxmm} which are in some cases a factor 5--10
lower than our previous estimated values 
(Tab.~\ref{tab:ul}) based on BAT data alone.
This in turn translates in larger intensities of the magnetic field
which in a few cases reach the $\sim0.5$\,$\mu$G value.

\section{Cluster Population}
\label{sec:pop}
%
%
\subsection{Cluster log N - log S}
\label{sec:logn}
Thanks to the serendipitous character of the BAT survey, it is possible
to derive, for the first time, the source counts distribution (also known
as log $N$ -- log $S$) of clusters above 15\,keV. This can be obtained as:
\begin{equation}
N(>S) = \sum_{i=0}^{N_S} \frac{1}{\Omega_i} \ \ \ \ \textup{[deg$^{-2}$]}
\end{equation}
where $N_S$ is the total number of clusters
with fluxes greater than $S$ and $\Omega_i$ 
is the geometrical area  surveyed to that limiting flux.
The cumulative distribution is reported in Fig.~\ref{fig:logn}.
Source counts distributions are generally fitted by a power law
of the form $N(>S) = A S^{-\alpha}$.
Given the small number of objects, we do not attempt a maximum likelihood
fit to derive the slope $\alpha$, but we note that
our flux distribution is consistent with an Euclidean function
$N \propto S^{-3/2}$ as shown in Fig.~\ref{fig:logn}. 
We derive the normalization $A$ as that one which
reproduces the number of observed objects above the flux of 
$\sim1\times10^{-11}$\,erg cm$^{-2}$ s$^{-1}$. 
Using the 90\,\% confidence limits for small numbers 
derived by \cite{gehrels86}, we find that a good representation
of our data is obtained by 
$N(>S)= (4.19^{+2.1}_{-1.4}\times 10^{-4}\textup{deg$^{-2}$}) S_{11}^{-1.5}$
where $S_{11}$ is the flux in unit of $10^{-11}$\,erg cm$^{-2}$ s$^{-1}$. 
This function is also shown in Fig.~\ref{fig:logn}.

Interestingly, we note that the integrated flux of all clusters
above $10^{-11}$\,erg cm$^{-2}$ s$^{-1}$ is 
9.7$\times10^{-11}$\,erg cm$^{-2}$ s$^{-1}$ sr$^{-1}$. This is only
$\sim$0.1\,\% of the Cosmic X-ray Background (CXB) flux as measured
by BAT in the 15--55\,keV band \citep{ajello08c}, but 5--10\,\% of 
the total flux resolved by BAT into AGNs \citep{ajello07b}.
Thus, clusters of galaxies are a sizable population among the extragalactic
objects (mostly AGN) detected by BAT.

We can compare the BAT log $N$ -- log $S$, with those derived 
in the 0.5--2\,keV band. In doing so, we extrapolate the BAT
spectra to the 0.5--2\,keV band using the temperatures measured
below 10\,keV. The cluster surface density above 
$10^{-12}$\,erg cm$^{-2}$ s$^{-1}$ in the 0.5--2.0\,keV band is 
4.3$^{+3.0}_{-2.3}\times10^{-2}$\,deg$^{-2}$ which is in rather
good agreement with the finding of \cite{vikhlinin98} and
\cite{burenin07}.

The BAT source counts distribution can be used to estimate the foreseen
number of galaxy clusters above a given flux limit.
In doing so, we adopt for $\alpha$ the -1.4 value which has been
established by deeper X-ray surveys
\cite[e.g.][and references therein]{jones98,boehringer01}.
Indeed, using the -3/2 value would certainly overestimate the cluster
density at lower fluxes.
As an example, an
instrument surveying the whole sky to $10^{-13}$\,erg cm$^{-2}$ s$^{-1}$
would detect approximately $\sim$10000 galaxy clusters in the 15--55\,keV
band. The BAT sample itself will comprise up to 30 objects,
if BAT will able to reach the 0.5\,mCrab flux limit on the whole sky.

%
%
\subsection{X--ray Luminosity Function}
Since all our clusters have a measured redshift, we can derive
their luminosity function. Its construction relies on the knowledge
of the survey volume $V_{max}$ as a function of X-ray luminosity.
The survey volume is the volume of the cone defined by the
survey area and the luminosity distance at which a cluster with a given
luminosity could just be observed at the flux limit.
The limiting luminosity distance $D_{L\ lim}$, and thus also $V_{max}$, 
can be determined
solving iteratively the following equation:
\begin{equation}
D^2_{L\ lim} = \frac{L_x}{4\pi F_{lim} k(T,z)}
\end{equation}
where $L_x$ is the source luminosity and 
$k(T,z)$ is the {\it k}-correction which accounts for the
redshifting of the source spectrum.

Once the $V_{max}$ is computed for each object, the cumulative
luminosity function can be derived as:
\begin{equation}
N(>L_x) = \sum_{i=0}^{N} \frac{1}{V_{max}(L_i)} \ \ \ \ 
\textup{[h$^{3}_{70}$\,Mpc$^{-3}$].}
\end{equation}

The cumulative luminosity function of the BAT clusters, obtained with
the method reported above, is shown in Fig.~\ref{fig:lumin}.
\cite{boehringer02}, analyzing a flux-limited sample of ROSAT galaxy
clusters (REFLEX), derived that a good parametrization
of the differential luminosity function is  
a Schechter function of the form:
\begin{equation}
\frac{dN}{dL} = n_0\ \textup{exp}\left(-\frac{L}{L_*}\right) 
\left(\frac{L}{L_*}\right)^{-\alpha} \frac{1}{L_*}.
\end{equation}
In order to compare the REFLEX luminosity function with the
BAT one, we adopt for $n_0$, $L_*$ and $\alpha$, the values determined
by \cite{boehringer02}. Moreover, since the REFLEX luminosity
function is derived in the 0.1--2.4\,keV band, we need to convert
the luminosities  to the BAT, 15--55\,keV, band. We do this
using the mean clusters temperature (kT = 8.1\,keV) 
determined in the 2--10\,keV band 
(see right panel of Fig.~\ref{fig:temp}).
The reason for adopting this  temperature instead of the
BAT-derived temperature is double. First, given the S/N ratio,
temperatures determined
in the 2--10\,keV band have a better accuracy than temperatures
determined in the BAT band. Most importantly however, using the 
2--10\,keV temperature allows a more accurate extrapolation of the
source luminosity from the ROSAT (0.1--2.4\,keV) to the BAT (15--55\,keV)
band. The extrapolated, cumulative, REFLEX luminosity function
is also  reported Fig.~\ref{fig:lumin}. It is apparent that, notwithstanding
the extrapolation, the 
agreement of the BAT data and the REFLEX luminosity function is excellent.
This agreement is not however surprising because most of  the BAT clusters
constitute the bright end of the REFLEX luminosity function.
The value of $L_*$ converted to the 15--55\,keV band is 
$L_*$=7.3$\times 10^{43}\ h^{-1}_{70}$\,erg s$^{-1}$ while
$n_0=5.13^{+2.7}_{-1.8}\times10^{-7}$ and $\alpha=1.63$

Integrating the luminosity function multiplied by the luminosity yields
the total X-ray emissivity $W$ of galaxy clusters. Above the survey
limit of $2\times 10^{43}$\,erg s$^{-1}$,  we find\footnotemark{}
\footnotetext{We do not provide an error estimate since the luminosity
function was not fitted to the data.} 
$W$=2.83$\times10^{37}$\,erg s$^{-1}$ Mpc$^{-3}$ (15--55\,keV). 
This can be compared to the 
total emissivity of AGN which was derived for the local Universe
and a similar energy band (17--60\,keV) by \cite{sazonov07}.
After correcting for the small difference between the  energy bands, 
the AGN local emissivity above $2\times 10^{43}$\,erg s$^{-1}$ 
is $W_{AGN}$=14.1$\times10^{37}$\,erg s$^{-1}$ Mpc$^{-3}$.
It is thus clear that galaxy clusters contribute substantially
($\sim$20\,\% level with respect to AGN) to the local X--ray output.

%
%
\section{Discussion}
\label{sec:disc}
\subsection{Non-thermal hard X-ray emission}
Direct evidence of the presence of relativistic electrons   in the ICM 
arises from the existence of large radio halos \citep{dennison80,feretti07}. 
The same electron
population responsible for the synchrotron emission can in principle
scatter CMB photons by IC and produce hard X-ray radiation. The intensity
of this radiation relative to the synchrotron emission
ultimately depends on the value of the magnetic field.

A firm detection of non-thermal components in the spectra of galaxy
clusters has remained elusive in the past as well as in this study.
Indeed, Perseus is the only galaxy cluster in the BAT sample 
where a non-thermal high-energy component is revealed 
at high significance. Most likely this component is 
due to the emission of the central AGN NGC 1275.
The rest of the clusters
detected by BAT do not show a significant non-thermal emission.
Using BAT data alone, we are able to constrain
 the non-thermal component below the mCrab level in the 
50--100\,keV energy band. 
The {\it BeppoSAX} detection above 50\,keV 
of an average non-thermal
component in the stacked spectrum of several clusters is consistent with
the BAT upper limit \citep{nevalainen04}.  
As discussed in $\S$~\ref{subsec:individual}, some of  the individual
detections of non-thermal components \citep[e.g.][]{eck07b} are consistent
(albeit some marginally) with the upper limits derived using BAT data alone.
Thus, we cannot exclude that such
non-thermal components exist and that they are currently below or 
at the limit of  the BAT sensitivity.  
If we assume that the principal emission mechanism 
is IC scattering of GeV electrons off CMB photons, then the cluster magnetic
field is constrained to be $\geq$0.1\,$\mu$G. These low magnetic intensities
would show
that the magnetic field is far from equipartition (i.e. the energy
in the magnetic field is different with respect to the electrons energy).
As pointed out by \cite{petrosian08}, this can happen if the sources
generating the magnetic field and accelerating the electrons are not identical.

However, IC emission by relativistic electrons can be modeled
as a power-law in the 1--200\,keV energy regime
\citep[e.g.][]{nevalainen04,reimer04} . Thus, 
using XMM-Newton/XRT/Chandra and  BAT data  we are able to constrain, more robustly,
the IC emission mechanism. With this approach, we confirm the detection
and the flux of the hard component in the spectrum of Abell 0754, but we 
are also able to prove (thanks to the resolution of XMM-Newton) that 
a single point-like object,  2MASS	09091372-0943047,
located less than 2\arcmin\ from the BAT centroid accounts for the whole
non-thermal emission.
For the rest
of the clusters, we are able to produce upper limits which are a factor
5--10 lower than previously estimated. These limits in turn translate
into a slightly larger intensity of the magnetic field  which
reduces the gap to Faraday rotation measurements \citep{kim91,clarke01}.
If the cluster magnetic field is truly of the $\mu$G order, then
the chances of detecting IC emission from clusters with
the currently flying instruments become really small \citep[][]{pfr04}.
Indeed, the values of the predicted IC flux
account for only $<$10\,\% of the claimed non-thermal X-ray emission
above 10\,keV when taking both primary and secondary-generated
electrons into account  \citep[see e.g.][]{miniati01}.
Recently, 
\cite{pfrommer07c},  using high-resolution simulations of a sample
of representative galaxy clusters, showed that the predicted
IC flux for the Coma and Perseus clusters would be a factor 50 lower than
the claimed detections.

Our combined analysis, thus, put tight constrains on the IC mechanism.
However, IC emission is the process that most likely explains the
claimed non-thermal emission, but not the only one.
Hard X-ray flux from galaxy clusters can be interpreted as bremsstrahlung
from supra-thermal electron tail developed in the thermal electron
distribution due to stochastic acceleration in the turbulent ICM
\citep[e.g.][]{ensslin99,petrosian01}. In this modeling, the radio
and the non-thermal X-ray flux are no longer strictly related and equipartition
may apply. However, the non-thermal bremsstrahlung model requires 
a continuos input of energy in the ICM which as a consequence
will cause its temperature to increase. Thus, the non-thermal bremsstrahlung
phase is likely to be short lived \citep{petrosian08}.
%
%

\subsection{Structure Formation}

All the galaxy clusters detected by BAT, except perhaps Abell 3571,
are merging systems. Some, as Abell 0754, Abell 2142 and, Abell 3266,
are experiencing violent merging due to encounters of subclusters
with comparable masses.
In the common scenarios of hierarchical structure formation 
\citep[e.g.][]{miniati00,ryu03}, large
systems evolve as the result of merging of smaller structures. 
As reviewed in \cite{dolag08}, cluster mergers generate internal
shocks (Mach number less than 4) which provide most of the ICM gas heating
\citep[e.g.][]{quilis98},
and also  likely convert a non-negligible fraction ($\leq$10\,\%)
 of their power into CRs.
The shocks primarily heat the ions because the kinetic energy of an ion
entering the shock region is larger than that of an electron by their
mass ratio \citep{takizawa99}.
Cosmological simulations have shown \citep[e.g.][]{pfrommer07},
that in the case of ongoing merger activity, the relative CR pressure
(to the thermal ICM pressure) is greatly enhanced, up to 15--20\,\%,
due to strong merger shock waves.  This pressure is likely larger
in the outskirts of the cluster because  of the lower sound 
speed and the larger density of the ICM in the central region which makes
CR acceleration less efficient \citep{pfrommer07}.

Hot spots as well as cold fronts have been found in many merging clusters 
thanks to the  superior resolution of Chandra
\citep[e.g.][]{markevitch00,markevitch01,markevitch03}. 
Hydrodynamical simulations have highlighted
that $\sim$1\,Gyr after the encounter of two clusters with comparable
masses post-shock regions with high temperatures (T $\approx$10--20\,keV ) 
are formed \citep[e.g.][]{takizawa99,richtie02}.
In the BAT sample there is a clear correlation of gas temperature and merging
activity. Indeed, Abell 3571 and Perseus, which are in a late merging 
stage, display the lowest plasma temperatures among the clusters in our sample.
INTEGRAL recently unveiled the presence of a hotter region (T$=12\pm2$\,keV)
located south-west of the centre of the Coma cluster \citep{eck07}.
These findings highlight the important role of merging shocks in the heating
of ICM.

\subsection{Clusters Statistics}
The serendipitous character of the BAT survey allowed us to
determine, for the first time above 10\,keV, the log {\it N} - log {\it S}
and luminosity function distributions of galaxy clusters.
Both are in very good agreement with previous studies. 
The log {\it N} - log {\it S} highlights that the clusters BAT detects
produce a negligible fraction ($\sim$0.1\,\%) of the X-ray background emission,
but they represent a sizable population (5--10\,\%) 
with respect to the local AGN.
The BAT log {\it N} - log {\it S} shows that future  instruments
with a sensitivity 10 or 100 times better than BAT (above 15\,keV) 
will detect clusters at a density of $\sim$0.01\,deg$^{-2}$ 
and  $\sim$0.24\,deg$^{-2}$ respectively.

The BAT luminosity distribution allowed us to determine that the volume
emissivity of galaxy clusters is 
$W(>2\times10^{43} \textup{erg s}^{-1})$=2.38$\times 10^{37}$\,erg s$^{-1}$ Mpc$^{-3}$ .
Above the same limiting luminosity, \cite{sazonov07} derived
that the volume emissivity of the local AGN is 
$W_{AGN}$=14.1$\times 10^{37}$\,erg s$^{-1}$ Mpc$^{-3}$. 
Thus, above $2\times10^{43}$\,erg s$^{-1}$, the cluster volume emissivity
is  20\,\% of that one of AGN.
Integrating the luminosity
functions to lower luminosity (e.g. $10^{41}$\,erg s$^{-1}$) changes
this fraction to $\sim$10\,\%. This change is due to the fact that
at low luminosity the AGN luminosity function is steeper than
the cluster luminosity function \citep[e.g.][]{sazonov07,boehringer02}.


\subsection{Future Prospects}
The study of non-thermal processes in clusters of galaxies requires a multi-wavelength approach. The ongoing {\it Swift}/BAT survey will likely
comprise up to 30 clusters if an all-sky sensitivity of 0.5\,mCrab
will be reached
and it will improve the signal-to-noise
ratio for the spectra of the clusters presented here.
Ultimately, major progresses are expected with the launch of 
Simbol-X\footnotemark{},
\footnotetext{http://www.asdc.asi.it/simbol-x/}
XEUS\footnotemark{},
\footnotetext{http://www.rssd.esa.int/index.php?project=xeus}
NUSTAR\footnotemark{},
\footnotetext{http://www.nustar.caltech.edu/} and 
NeXT\footnotemark{}.
\footnotetext{http://www.astro.isas.ac.jp/future/NeXT/}
Indeed, their sensitivities and spectro-imaging capabilities up to 
high energies ($80$ keV and beyond) will provide new and better 
constraints on the hard X-ray emission.

The future generation of radio arrays combined with high-energy  observations will allow to shed some light on the
energetics of relativistic  particles, the nature and frequency of acceleration processes, and the strength and structure of magnetic fields. As we already discussed, this astrophysical information has strong cosmological implications.
The Long Wavelength Array\footnotemark{}
\footnotetext{http://lwa.unm.edu} (LWA),
the Low Frequency Array\footnotemark{} (LOFAR),
\footnotetext{http://www.lofar.org/}
and ultimately the Square Kilometre Array\footnotemark{} 
\footnotetext{http://www.skatelescope.org} (SKA), 
will operate over a critical radio frequency range to detect relativistic
plasma in large-scale structure and clusters in a sensitive way.
The advance in sensitivity and resolution will increase the statistics of known radio halos and radio relics at different 
 redshifts. The correlation of sensitive X-ray and radio detections
 will be particularly important \citep[e.g.][]{ensslin02}. 
At the same time,  thanks to the high angular and spectral resolution,
the Faraday rotation studies will significantly 
improve yielding a better determination of the cluster magnetic field.

Much attention is directed towards the Gamma-ray Large Area Space Telescope
\footnotemark{}\footnotetext{http://www-glast.stanford.edu}(GLAST) which,
with an unprecedented sensitivity, spatial resolution and dynamic range
at GeV energies, will shed light
on the origin of the extragalactic $\gamma$-ray background.
Galaxy clusters and shocks from structure formations are natural
candidates for explaining part of this diffuse emission 
\citep[e.g.][and references therein]{dermer07}. All the BAT clusters
are good candidate for GLAST since they are nearby and are mergers.
Indeed, in merging systems, part of the internal shocks energy is very
likely converted into CRs acceleration \citep{dolag08}. 
As pointed out by \cite{pfrommer07_2}, above 100\,MeV the cluster
emission will likely be dominated by pion decay $\gamma$-rays even
though a contribution from non-thermal bremsstrahlung and IC emission
of secondary electrons is expected.
This will provide a unique information about the hadron component of CRs which
is not included in estimates of CR pressure based only on the
observations discussed above concerning electrons and magnetic field.
Since cosmic ray protons loss time is long, 
the $\pi^0$-bump detection would proof that hadrons
have  been confined in the ICM for as 
long as the Hubble time  \citep[e.g.][]{berezinsky97}. 
Stringent constraints 
on the cosmic rays content in the ICM is fundamental for the future  
space missions which will use galaxy clusters to constrain and understand
the nature of Dark Energy 
(e.g. eROSITA\footnotemark{}\footnotetext{http://www.mpe.mpg.de/projects.html\#erosita}).

%
%
\section{Conclusions}
\label{sec:concl}
BAT is the first instrument to detect above 15\,keV 
an all-sky sample of galaxy clusters\footnotemark{}. 
\footnotetext{We are aware of an independent work \citep{okajima08} 
based on an alternative analysis of BAT survey data which 
reaches conclusions consistent with this analysis.}
The BAT energy range (15--200\,keV) 
is the best one to investigate the presence of non-thermal emission
whose detection   remained so far controversial.
The results of our investigation can be summarized as follows:
\begin{itemize}
\item Perseus is the only cluster among the 10 BAT objects which
displays an high-energy non-thermal component which extends up to 200\,keV.
It is very likely that the central AGN NGC 1275 is responsible for 
such emission. 
This claim is supported by several evidences: 
1) the  variability seen with BeppoSAX \citep{nevalainen04}, 
2) the XMM-Newton spectral analysis \citep{churazov2003},
and 3) our combined BAT--XRT--XMM-Newton
 analysis which shows that the nucleus has
a typical AGN spectrum.

\item The BAT spectra of the remaining
9 galaxy clusters is well fitted by a simple thermal model that constrains
non-thermal flux to be below 1\,mCrab in the 50--100\,keV band.
\item Assuming that IC scattering is the main mechanism at work for 
producing non-thermal high-energy flux, it is possible to estimate
the magnetic field using Radio data and 
the upper limits derived above. 
We obtain that  in all the BAT clusters the (average) 
magnetic field is $>0.1$\,$\mu$G. These (rather uncertain) values are 
in disagreement (if the magnatic field 
intensities are close to the lower limits)
with the, also uncertain, 
Faraday rotation measurements which show that the magnetic field is in the
$\sim$\,$\mu$G range. Our low magnetic field values would imply that the 
magnetic field is far from equipartition.
\item The stacked spectrum of the BAT clusters 
(except Perseus and Coma) confirms
once again the absence of any non-thermal high-energy component. 
The $\sim$56\,Ms stacked spectrum constrains any non-thermal flux 
to be below 0.3\,mCrab (or 1.9$\times 10^{-12}$\,erg cm$^{-2}$s$^{-1}$)
in the 50--100\,keV band.
\item  Using Swift/XRT, XMM-Newton and Chandra, 
in addition to BAT data, we were able
to produce X--ray cluster spectra which extend more than 3 decades in energy
(0.5--50\,keV). In all cases, but Perseus and Abell 0754, the broad-band
X--ray spectrum is well approximated by a single-temperature thermal model.
These spectra allowed us to put constrains on the IC emission mechanism which
are a factor $>$5 lower than those derived using BAT data alone. This would
in turn imply a larger intensity of the magnetic field.
For both Perseus and Abell 0754 an additional power-law 
component is statistically 
required, but several evidences confirm that two X-ray point sources 
(NGC 1275 and  2MASS	09091372-0943047) account for the total non-thermal
emission.
\item The cluster centroid shift in different wavebands,
the morphology and the complex temperature maps (available in literature),
show that 8 out of 10 clusters are in the middle of a
major merging phase. Shocks, which are revealed by XMM-Newton and Chandra 
images,
are actively heating the ICM as the BAT high temperatures testify.
The BAT observations and limits on the non-thermal emissions
can help to calibrate the large scale structure
formation simulations focusing in particular on the treatment of
non-thermal particle emission and cooling.
\item We have produced the first cluster source count (also
known as log {\it N} - log {\it S}) distribution above 15\,keV. 
This shows that, at the limiting fluxes sampled by BAT,
the surface density of clusters is $\sim$5\,\% of that one of AGNs.
Moreover, we find that the contribution of clusters to the Cosmic
X-ray background is of the $\sim$0.1\,\% order in the 15--55\,keV band.
The BAT log {\it N} - log {\it S} can be used to predict the cluster
surface density for future hard X-ray instruments.
\item The X-ray luminosity function of the BAT clusters, the first derived
above 15\,keV, is in excellent agreement with the ROSAT 
luminosity function derived in the 0.1--2.4\,keV band.
\end{itemize}

\acknowledgments
MA acknowledges funding from the DFG Leibniz-Prize (HA 1850/28-1).
PR is supported by the Pappalardo Postdoctoral Fellowship in Physics at MIT.
NC was partially supported from a NASA grant NNX07AV03G.
MA and PR wish to acknowledge Bal\'{u} for his incomparable constant enthusiasm.
We thank T. Okajima for providing a copy of his manuscript before submission
and for interesting discussions. 
The anonymous referee is aknowledged for his helpful comments which
improved the manuscript.
This research has made use of the NASA/IPAC extragalactic Database (NED) which
is operated by the Jet Propulsion Laboratory, of data obtained from the 
High Energy Astrophysics Science Archive Research Center (HEASARC) provided 
by NASA's Goddard Space Flight Center, of the SIMBAD Astronomical Database
which is operated by the Centre de Donn\'ees astronomiques de Strasbourg,
and of the ROSAT All Sky
Survey maintained by the Max Planck Institut f\"ur Extraterrestrische Physik. 

{\it Facilities:} \facility{Swift (BAT/XRT)}, \facility{XMM-Newton,} 
\facility{Chandra} .

\bibliographystyle{apj}
\bibliography{/Users/marcoajello/Work/Papers/Clusters/biblio}



\begin{figure*} 
    \begin{center}
     \begin{tabular}{c}
  \includegraphics[scale=0.35,angle=270]{f7a.ps} \\
  \includegraphics[scale=0.35,angle=270]{f7b.ps} \\
  \includegraphics[scale=0.35,angle=270]{f7c.ps} \\
     \end{tabular}
    \end{center}
    \null\vspace{-7mm}
    \caption{Residuals to the fit to Coma data using: 
a single thermal model (top),
sum of a thermal model and a power law (middle),
and the sum of two thermal models (bottom).
}
  \label{fig:rescoma}
\end{figure*}

\begin{figure*} 
    \begin{center}
     \begin{tabular}{cc}
    \includegraphics[scale=0.4,angle=0]{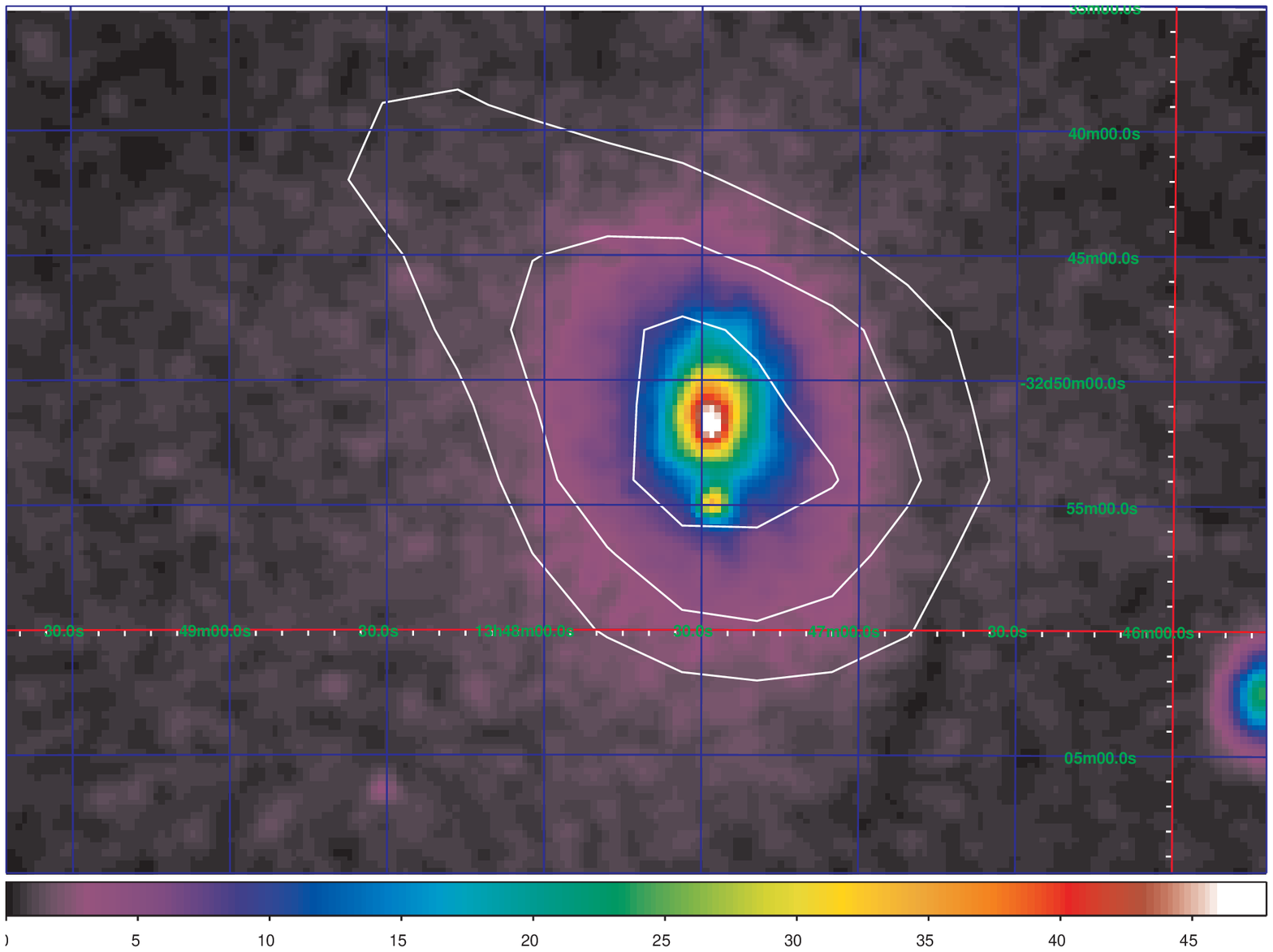} &
   \includegraphics[scale=0.45,angle=0]{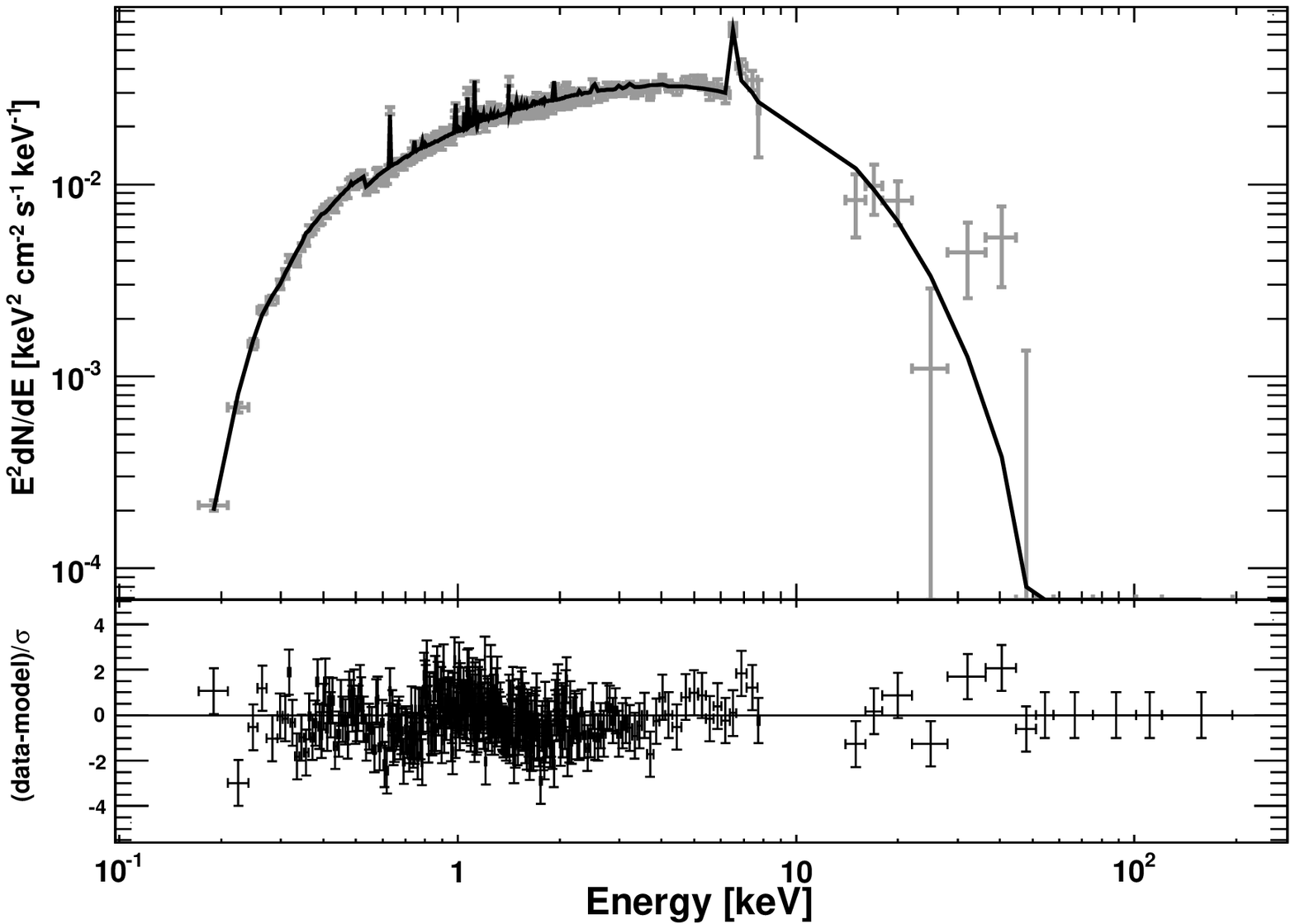} \\
    \end{tabular}
    \end{center}
    \null\vspace{-7mm}
   \caption{
{\it Left Panel:} ROSAT 0.1--2.4\,keV surface brightness of Abell 3571
with BAT significance contours superimposed. The contours range from 
2.5\,$\sigma$ to 5.0\,$\sigma$.
{\it Right Panel:} Joint fit to XMM-Newton--BAT data with 
a thermal model. The best fit model is shown as a solid line.
}
  \label{fig:abell3571}
\end{figure*}	
\begin{figure*} 
    \begin{center}
     \begin{tabular}{cc}
    \includegraphics[scale=0.4,angle=0]{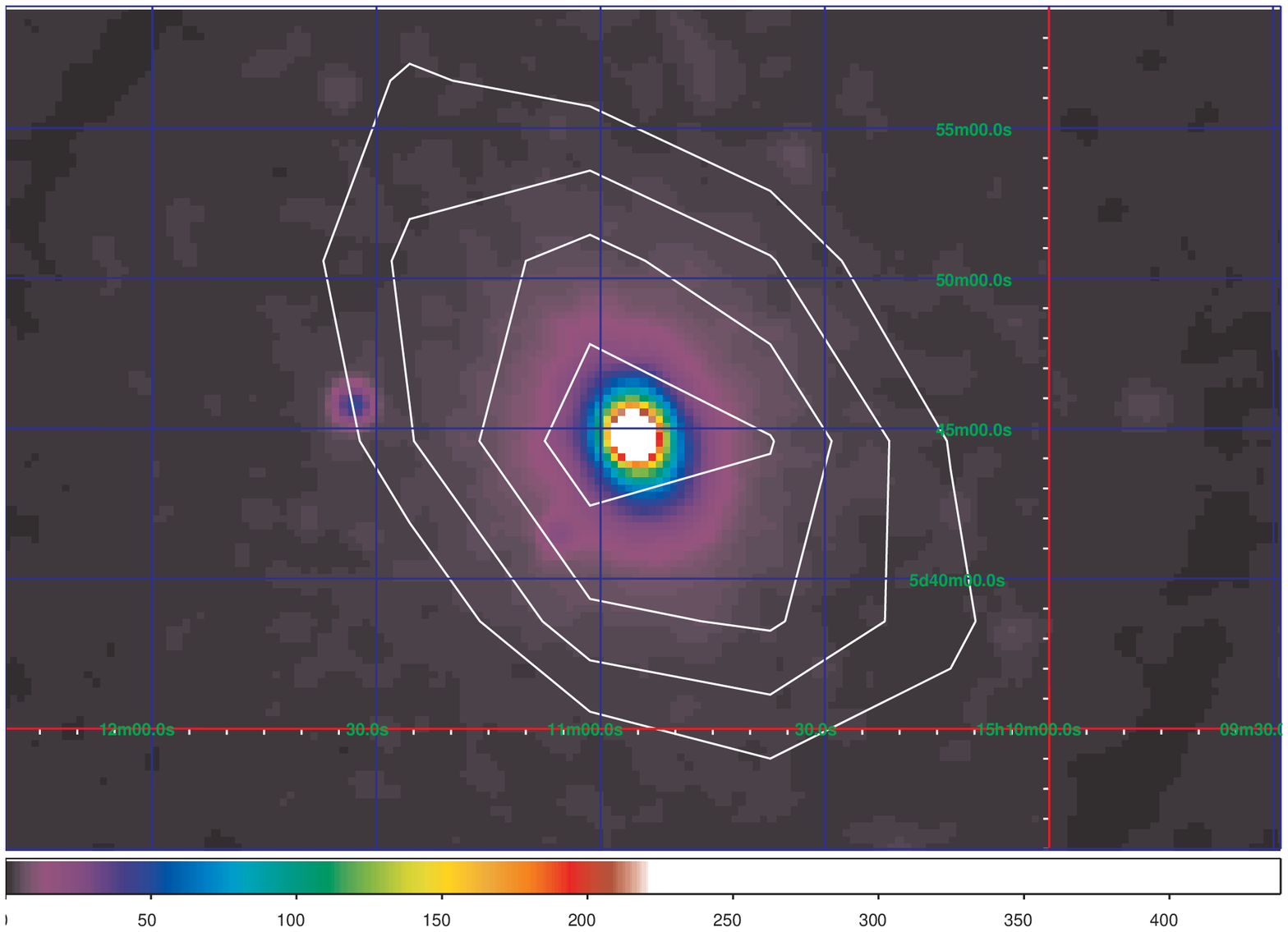} &
   \includegraphics[scale=0.45,angle=0]{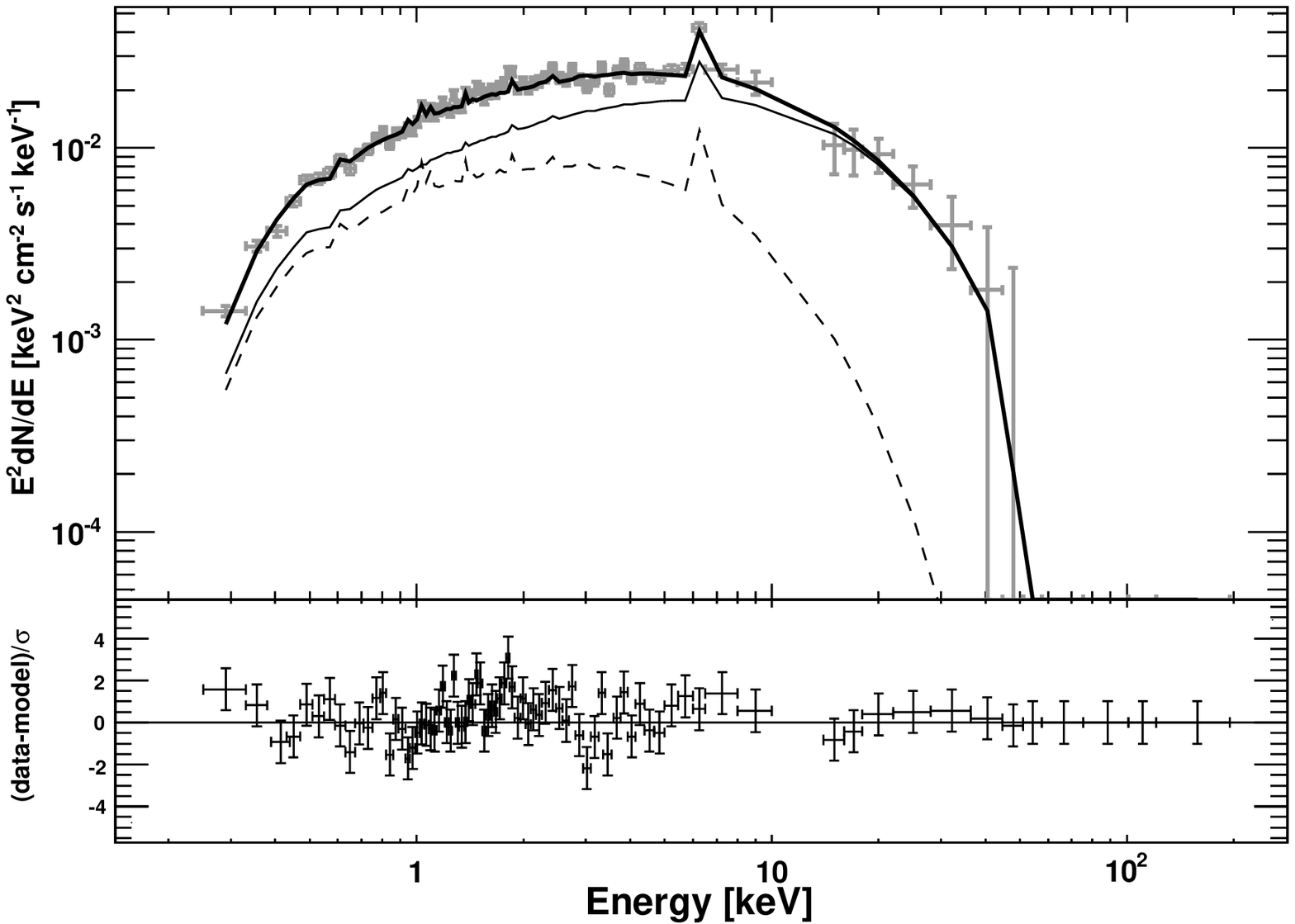} \\
    \end{tabular}
    \end{center}
    \null\vspace{-7mm}
   \caption{
{\it Left Panel:} ROSAT 0.1--2.4\,keV surface brightness of Abell 2029
with BAT significance contours superimposed. The contours range from 
2.5\,$\sigma$ to 5.0\,$\sigma$.
{\it Right Panel:} Joint XRT--BAT spectrum of Abell 2029. The best fit 
(thick solid line) is the sum of two thermal models 
(thin solid and dashed line).
}
  \label{fig:abell2029}
\end{figure*}	
\begin{figure*} 
    \begin{center}
     \begin{tabular}{cc}
    \includegraphics[scale=0.4,angle=0]{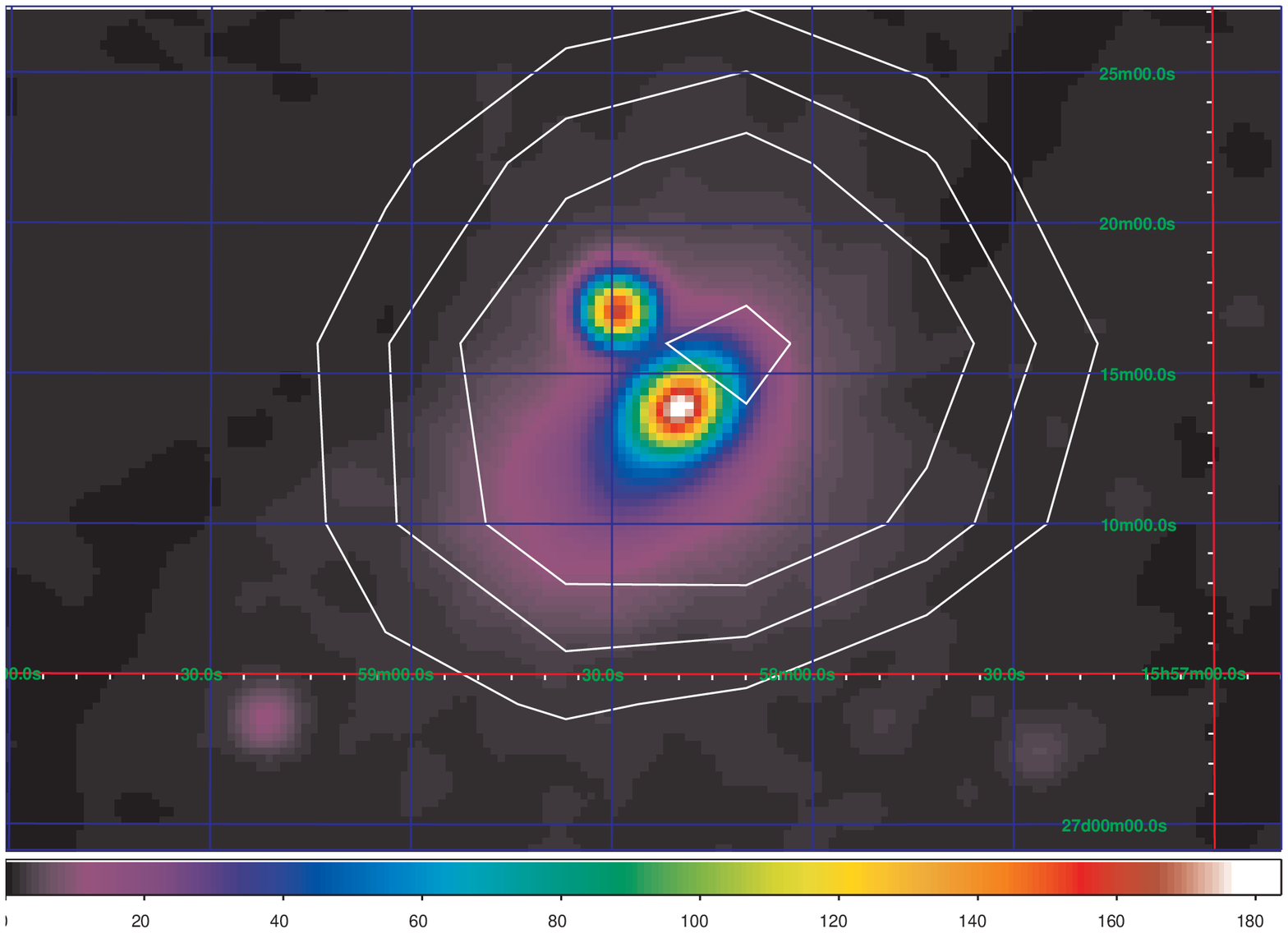} &
   \includegraphics[scale=0.45,angle=0]{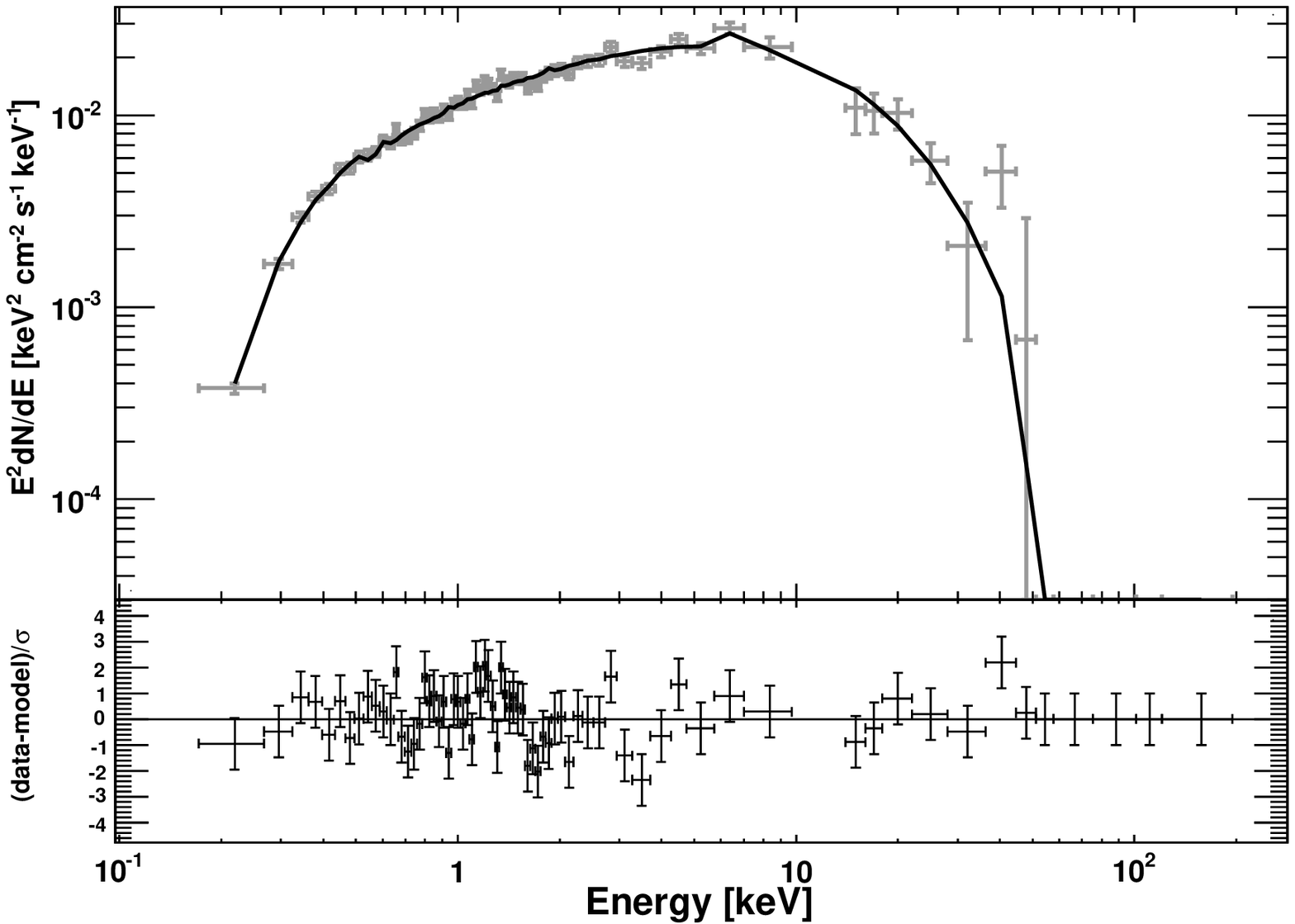} \\
    \end{tabular}
    \end{center}
    \null\vspace{-7mm}
   \caption{
{\it Left Panel:} ROSAT 0.1--2.4\,keV surface brightness of Abell 2142
with BAT significance contours superimposed. The contours range from 
2.5\,$\sigma$ to 7.0\,$\sigma$.
{\it Right Panel:} Joint fit to XMM-Newton--BAT data for Abell 2142
with single thermal model.
The best fit model is shown as a solid line.
}
  \label{fig:abell2142}
\end{figure*}	
\begin{figure*} 
    \begin{center}
     \begin{tabular}{cc}
    \includegraphics[scale=0.3,angle=0]{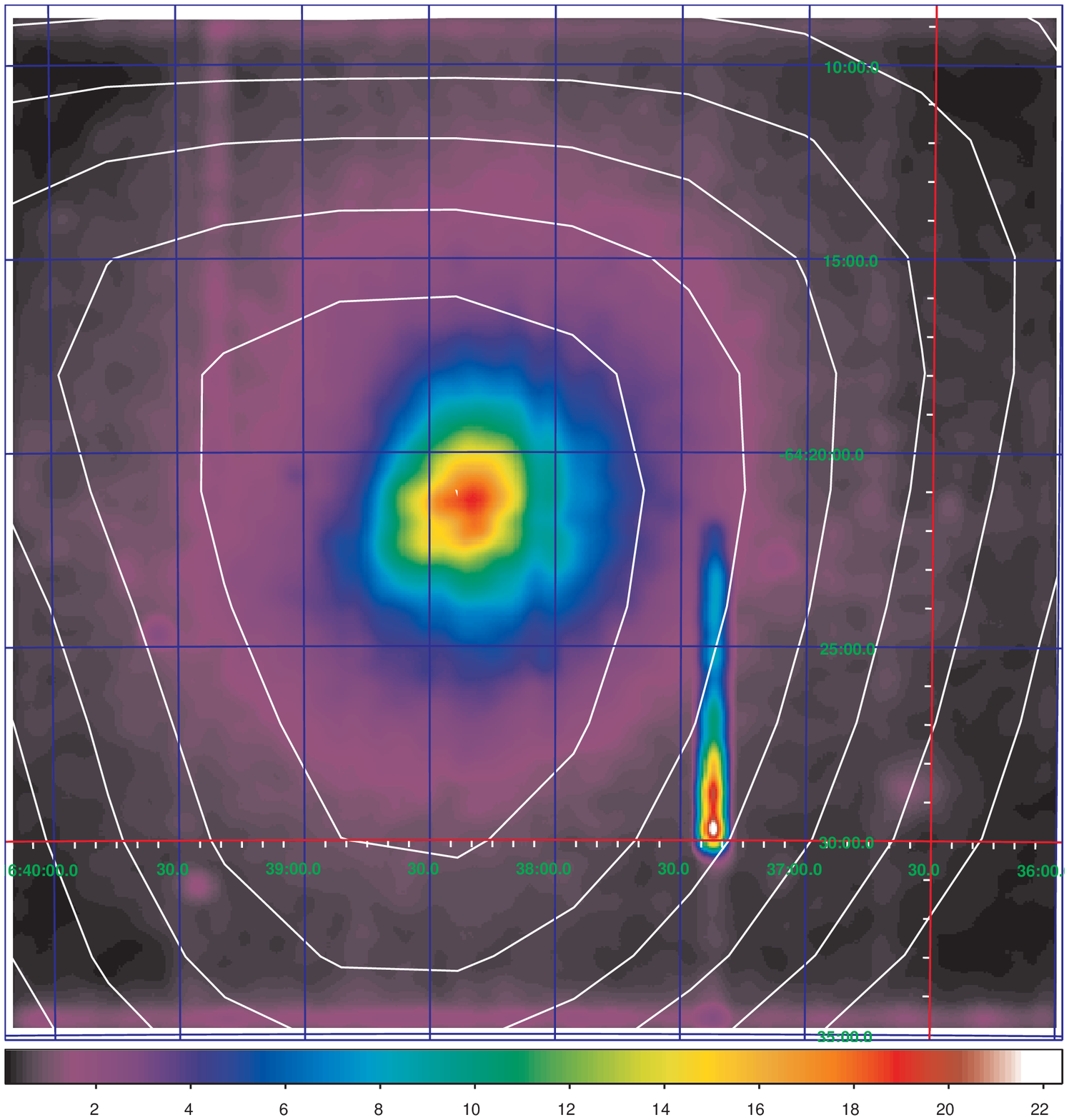} &
   \includegraphics[scale=0.45,angle=0]{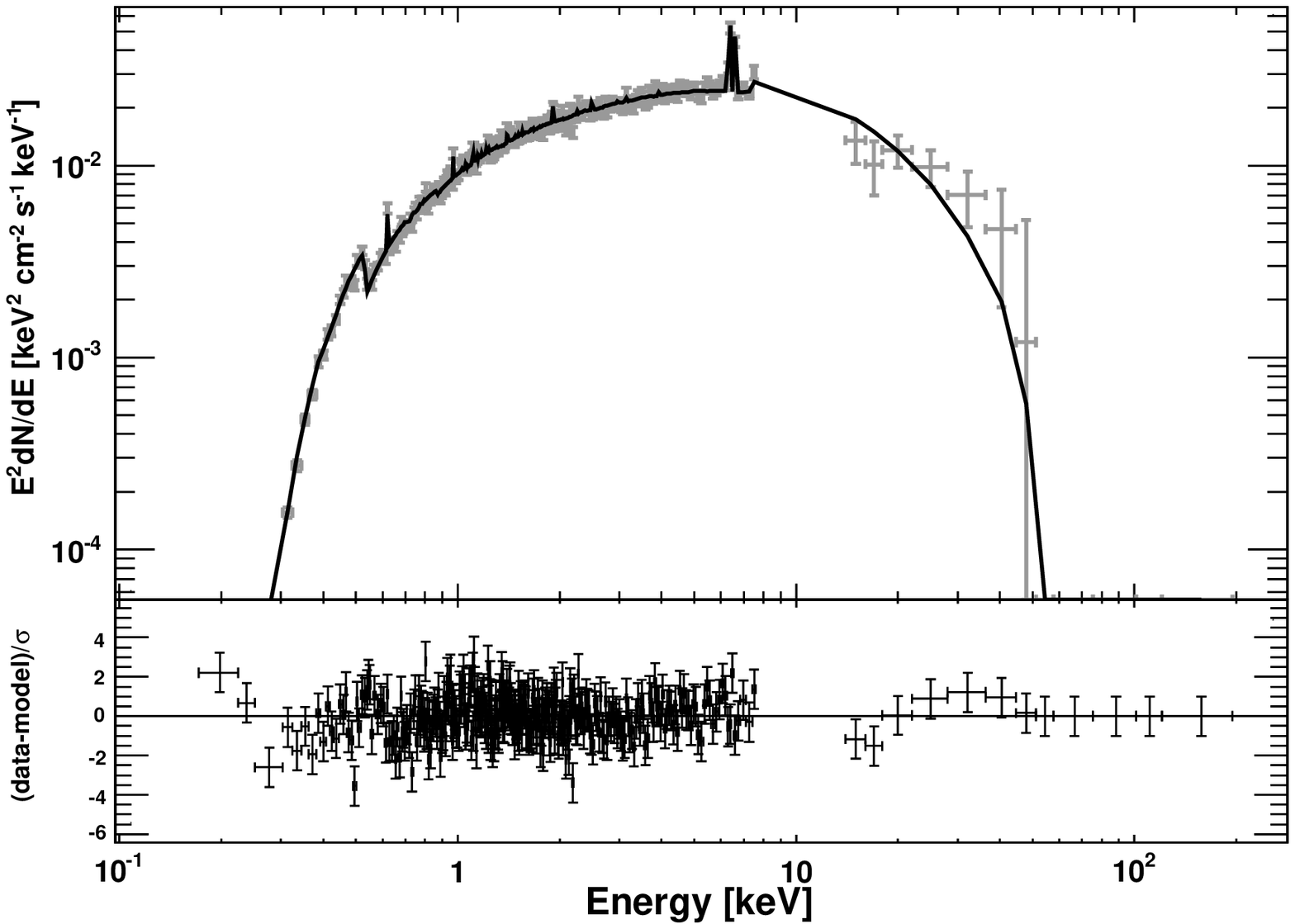} \\
    \end{tabular}
    \end{center}
    \null\vspace{-7mm}
   \caption{
{\it Left Panel:} XMM-Newton 1.0--7.0\,keV 
surface brightness of the Triangulum
Australis cluster
with BAT significance contours superimposed. The contours range from 
2.5\,$\sigma$ to 7.0\,$\sigma$.
{\it Right Panel:}
Joint fit to XMM-Newton--BAT data for the Triangulum Australis cluster
with a  thermal model.
The best fit model is shown as a solid line.
}
  \label{fig:triangulum}
\end{figure*}	
\begin{figure*} 
    \begin{center}
     \begin{tabular}{cc}
    \includegraphics[scale=0.4,angle=0]{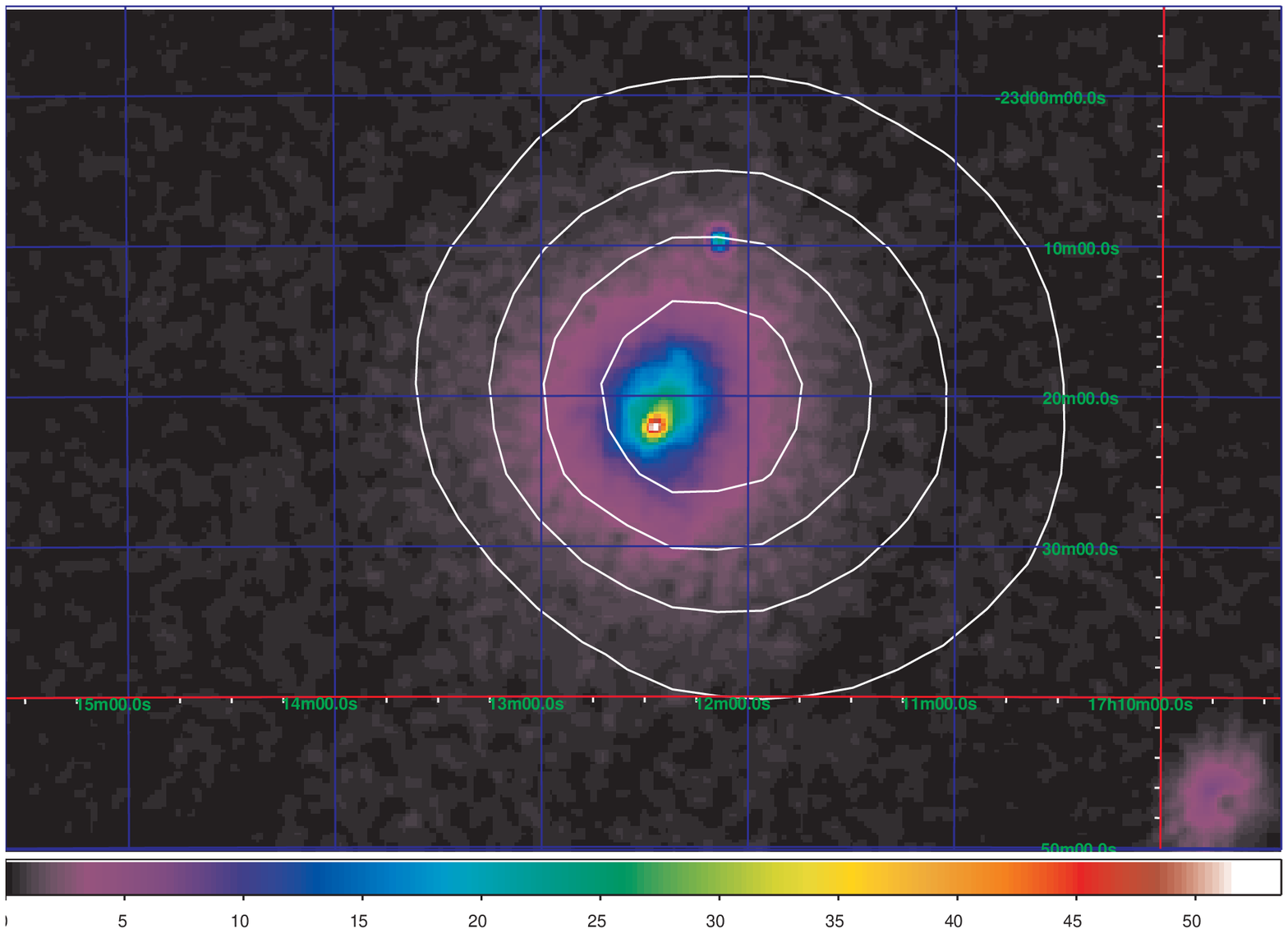} &
   \includegraphics[scale=0.45,angle=0]{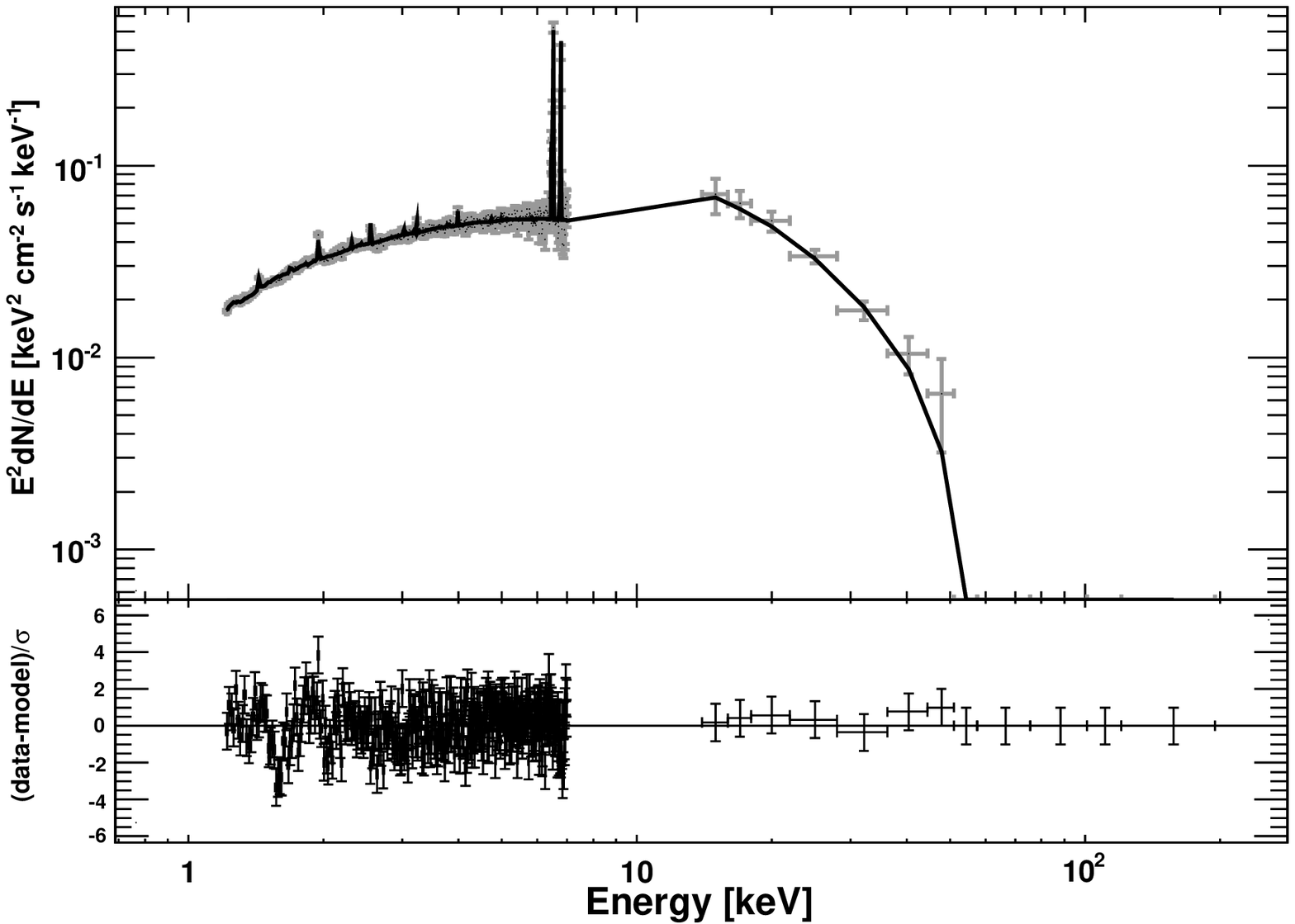} \\
    \end{tabular}
    \end{center}
    \null\vspace{-7mm}
   \caption{
{\it Left Panel:} ROSAT 0.1--2.4\,keV 
surface brightness of the Ophiucus cluster
with BAT significance contours superimposed. The contours range from 
2.5\,$\sigma$ to 22\,$\sigma$.
{\it Right Panel:} Joint fit to Chandra--BAT data for the 
Ophiucus cluster with a thermal model. The best fit is shown as solid line.
}
  \label{fig:ophiucus}
\end{figure*}

\begin{figure*} 
    \begin{center}
     \begin{tabular}{cc}
    \includegraphics[scale=0.4,angle=0]{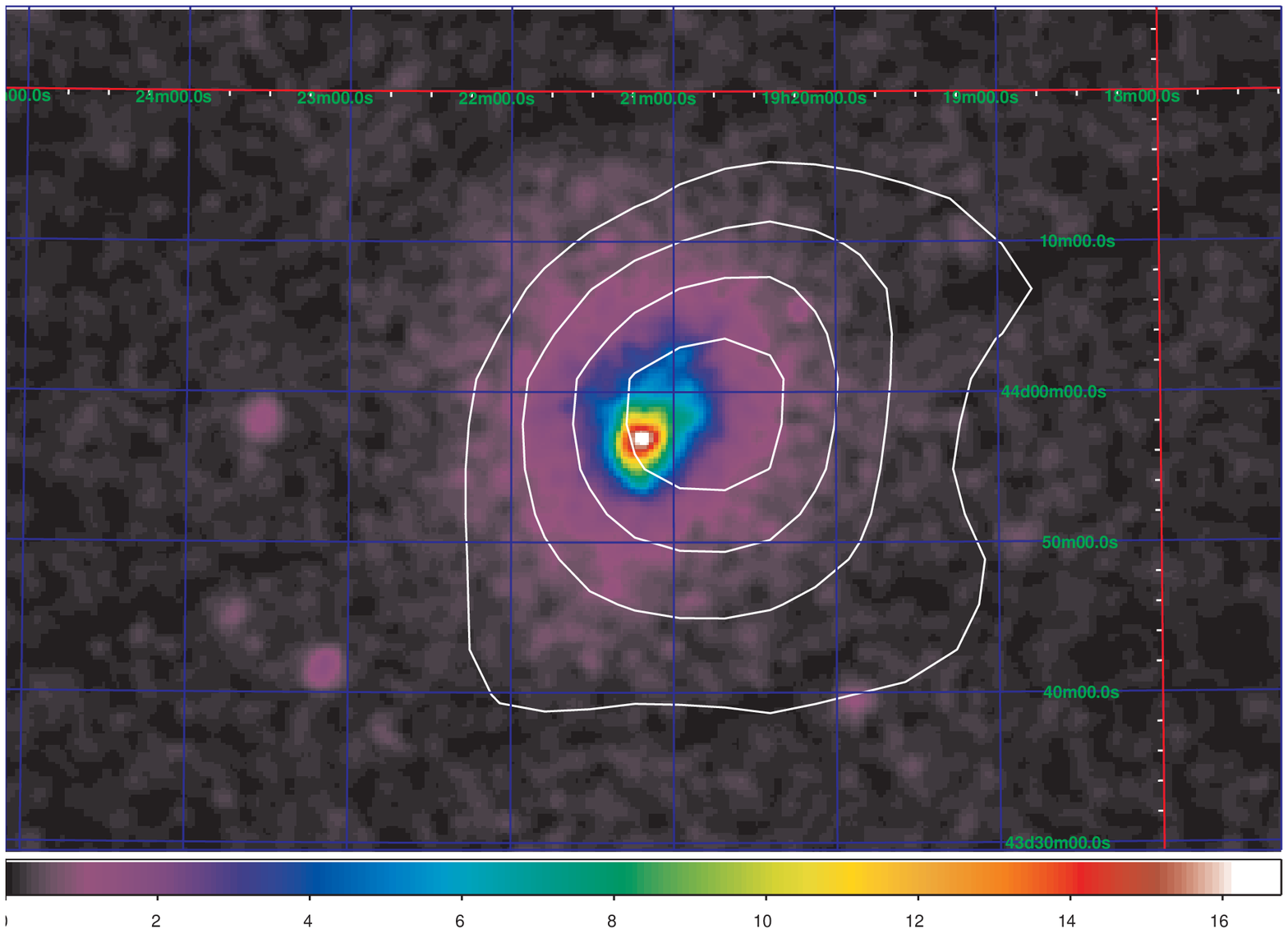} &
   \includegraphics[scale=0.45,angle=0]{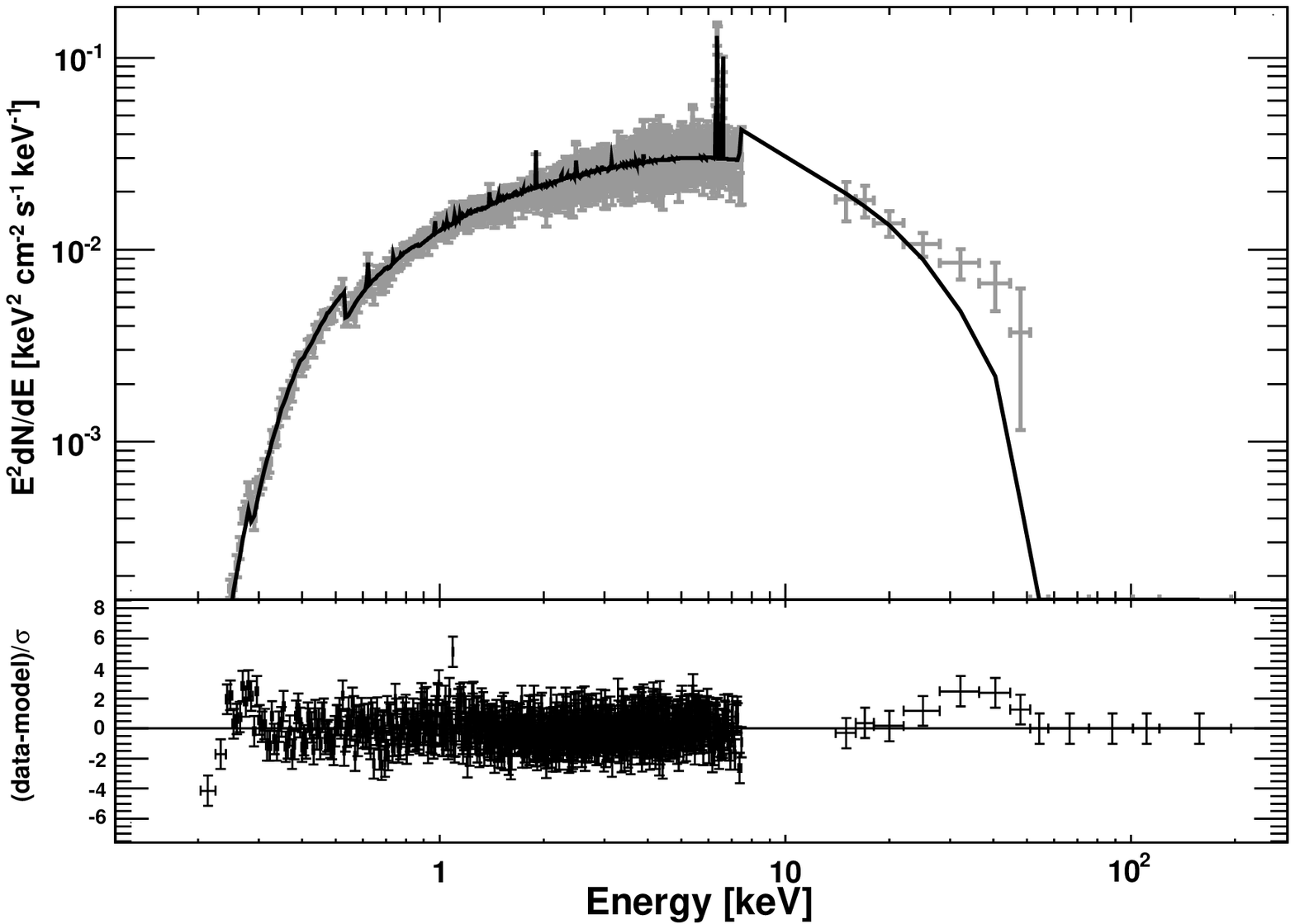} \\
    \end{tabular}
    \end{center}
    \null\vspace{-7mm}
   \caption{
{\it Left Panel:} ROSAT 0.1--2.4\,keV 
surface brightness of Abell 2319
with BAT significance contours superimposed. The contours range from 
2.5\,$\sigma$ to 22\,$\sigma$.
{\it Right Panel}
Joint fit to XMM-Newton--BAT data for Abell 2319 Australis cluster.
The best fit model  thermal model is shown as a solid line.
}
  \label{fig:abell2319}
\end{figure*}

\clearpage
\begin{figure}
\begin{center}
\includegraphics[scale=0.6]{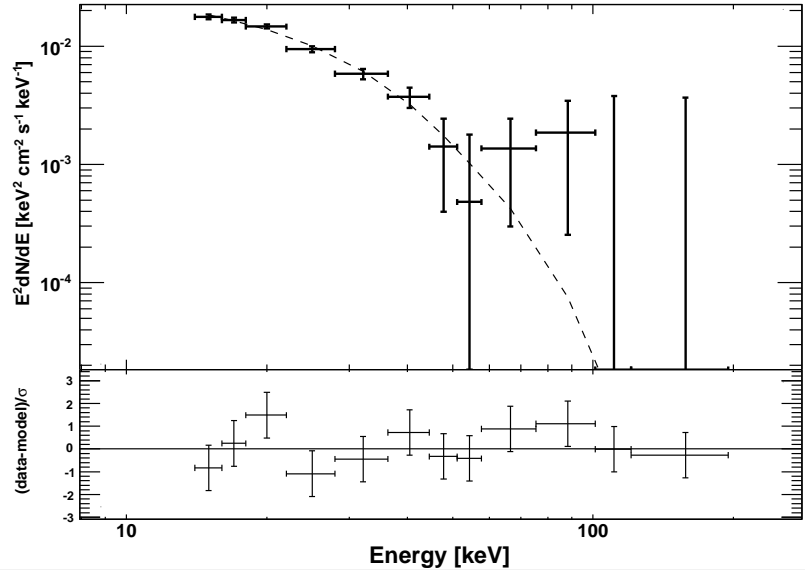}
\caption{
Stacked spectrum of the clusters in our sample
and the best fit (dashed line) with a bremsstrahlung model.
\label{fig:stacked}}
\end{center}
\end{figure}

\begin{figure}
\begin{center}
\includegraphics[scale=0.6]{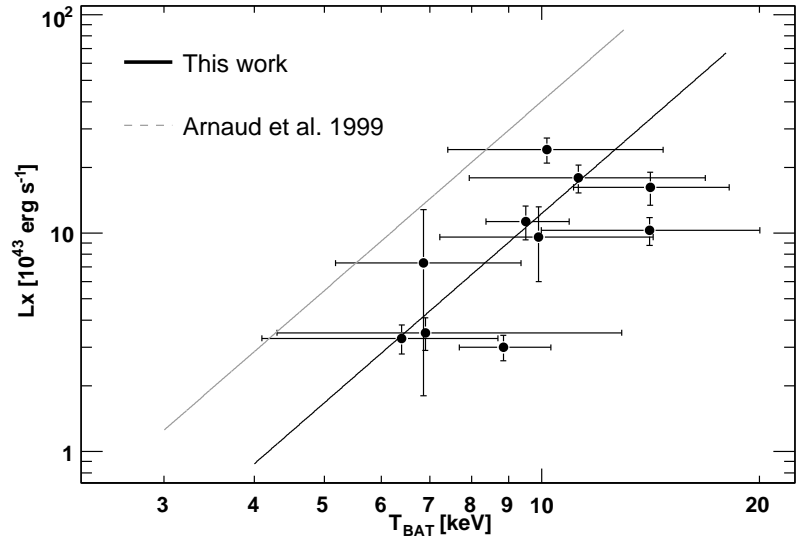}
\caption{Luminosity-Temperature relation for the BAT clusters.
The black line is the best, power-law, fit to the data while
 the gray line is the best fit of \cite{arnaud99} converted to the BAT
energy band.
\label{fig:lx_vs_T}}
\end{center}
\end{figure}

\begin{figure}
\begin{center}
\includegraphics[scale=0.6]{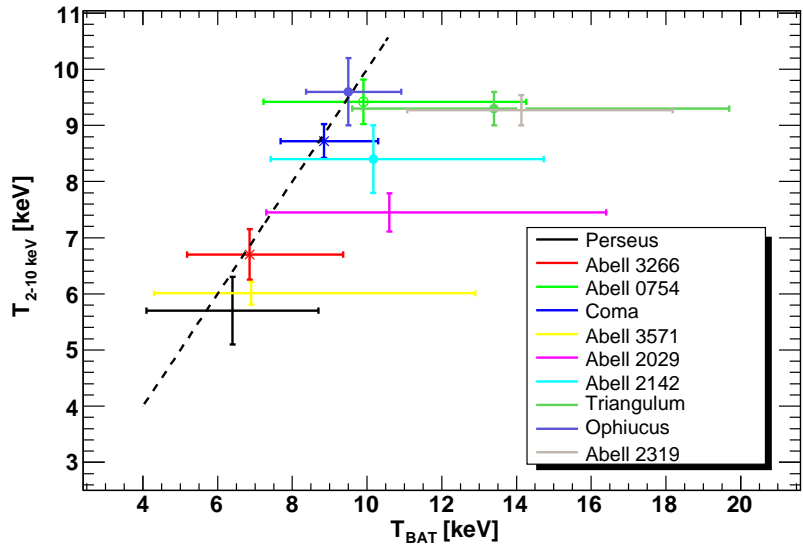}
\caption{
Comparison of best fit gas temperatures. The x-axis
reports the temperatures derived using BAT data (above 15\,keV) while
the y-axis shows the temperatures derived using 2--10\,keV data (XMM-Newton, Chandra
or XRT). The dashed line shows the $T_{BAT}=T_{2-10\,keV}$ function.
The largest deviations are for the merging clusters Abell 2029 and Abell 2319.
\label{fig:temp}}
\end{center}
\end{figure}

\begin{figure*} 
    \begin{center}
     \begin{tabular}{c}
  \includegraphics[scale=0.35,angle=270]{f17a.ps} \\
  \includegraphics[scale=0.35,angle=270]{f17b.ps} \\
  \includegraphics[scale=0.35,angle=270]{f17c.ps} \\
     \end{tabular}
    \end{center}
    \null\vspace{-7mm}
    \caption{Residuals to the fit to Abell 2029 data using: 
a single thermal model (top),
sum of a thermal model and a power law (middle),
and the sum of two thermal models (bottom).
}
  \label{fig:res2}
\end{figure*}	

\begin{figure*}
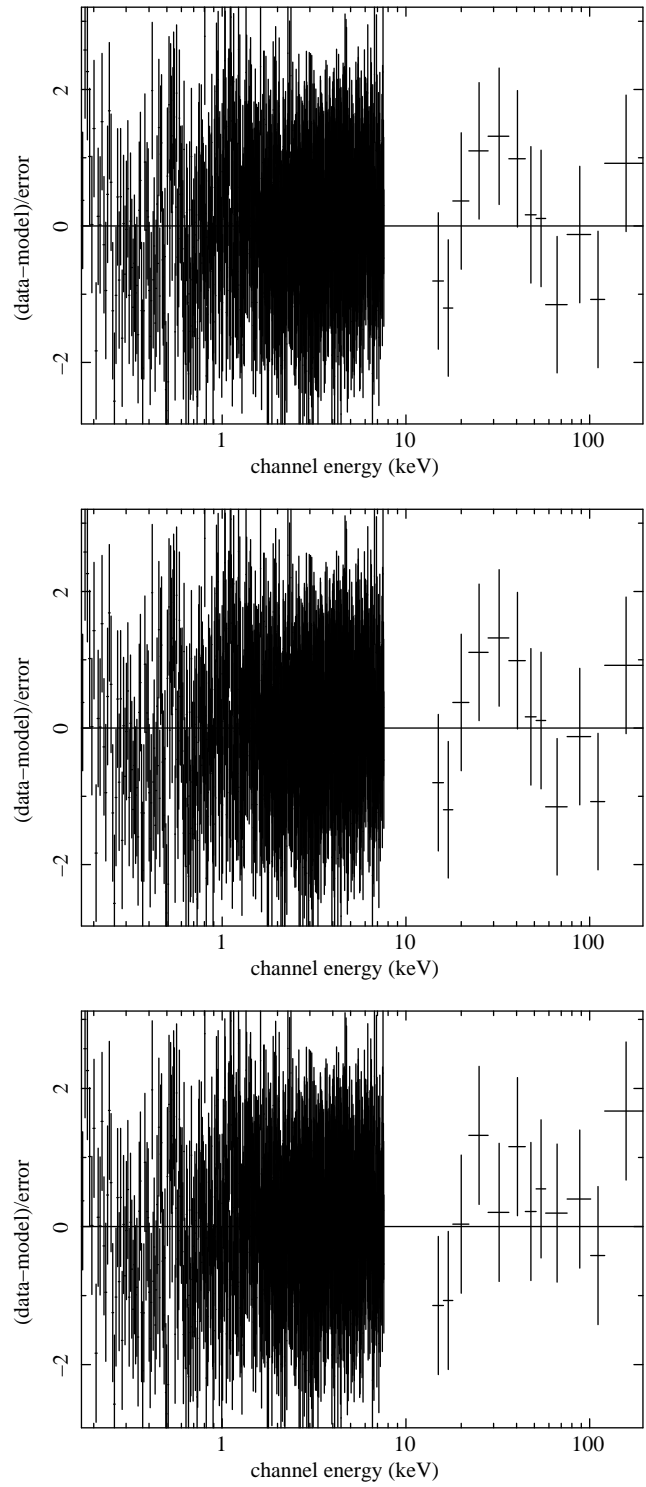
 
    \begin{center}
     \begin{tabular}{c}
  \includegraphics[scale=0.35,angle=270]{f18a.ps} \\
  \includegraphics[scale=0.35,angle=270]{f18b.ps} \\
  \includegraphics[scale=0.35,angle=270]{f18c.ps} \\
     \end{tabular}
    \end{center}
    \null\vspace{-7mm}
    \caption{Residuals to the fit to Triangulum Australis data using: 
a single thermal model (top),
sum of a thermal model and a power law (middle),
and the sum of two thermal models (bottom).
}
  \label{fig:res3}
\end{figure*}	

\begin{figure*} 
    \begin{center}
     \begin{tabular}{c}
  \includegraphics[scale=0.35,angle=270]{f19a.ps} \\
  \includegraphics[scale=0.35,angle=270]{f19b.ps} \\
  \includegraphics[scale=0.35,angle=270]{f19c.ps} \\
     \end{tabular}
    \end{center}
    \null\vspace{-7mm}
    \caption{Residuals to the fit to Abell 2319 data using: 
a single thermal model (top),
sum of a thermal model and a power law (middle),
and the sum of two thermal models (bottom).
}
  \label{fig:res4}
\end{figure*}

\clearpage
\begin{figure}
\begin{center}
\includegraphics[scale=0.6]{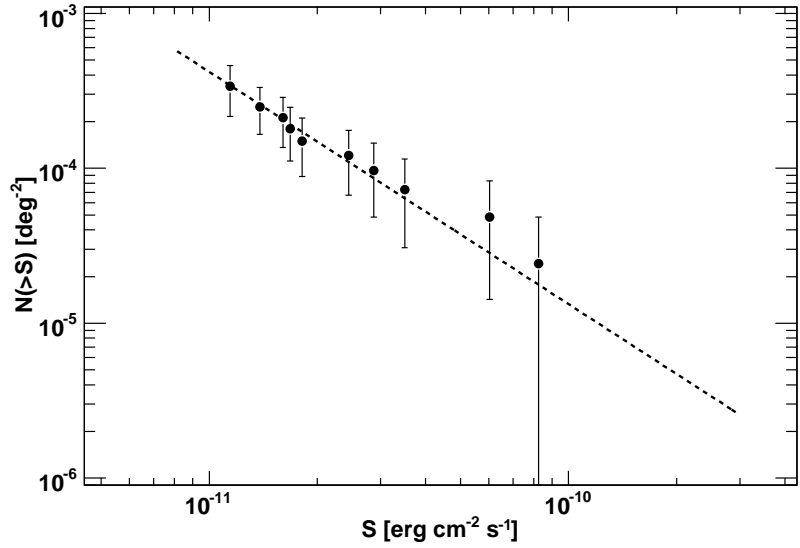}
\caption{Cumulative flux number relation for the BAT clusters (15--55\,keV).
The dashed line is an overlaid power law N($>$S) = A S$^{-1.5}$.
\label{fig:logn}}
\end{center}
\end{figure}

\begin{figure}
\begin{center}
\includegraphics[scale=0.6]{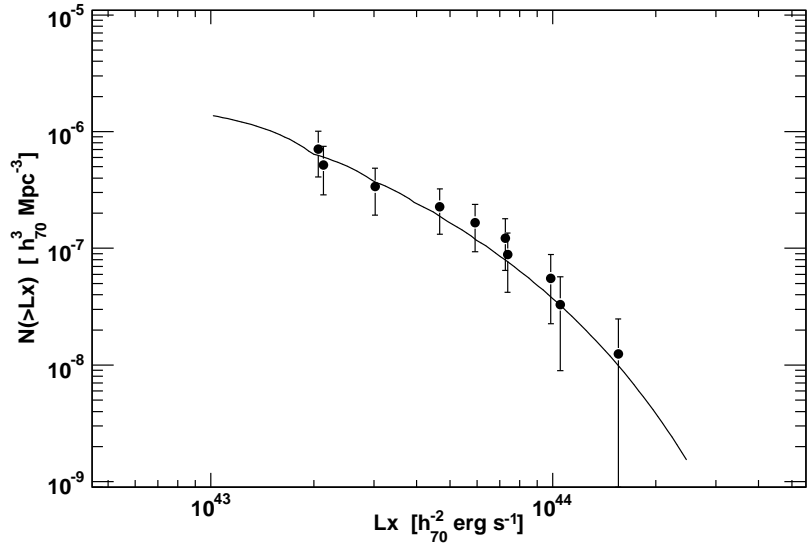}
\caption{Cumulative luminosity function of the BAT clusters (15--55\,keV).
The solid line is the X-ray luminosity function determined
for the REFLEX survey \citep{boehringer02}  converted to the BAT energy
band. 
\label{fig:lumin}}
\end{center}
\end{figure}

\begin{deluxetable}{lccclccc}
\tablewidth{0pt}
\tablecaption{Clusters detected in the 15--55 keV band \label{tab:clusters}}
\tablehead{
\colhead{NAME}               & \colhead{R.A.}       &
\colhead{DEC}            &  \colhead{S/N}      & \colhead{ID}  & \colhead{z}
& \colhead{EXPOSURE}     & \colhead{OFFSET} \\
\colhead{}               & \colhead{\scriptsize (J2000)}  &
\colhead{\scriptsize (J2000)}      & & &
& \colhead{\scriptsize{(Ms) }}     & \colhead{\scriptsize(arcmin)}
}
\startdata
J0319.8+4130 &   49.9573  &  41.5110   &   28.00 & Perseus        & 0.0175 & 2.89 & 0.5    \\
J0431.3-6126 &   67.8297  & -61.4388   &    5.61 & Abell 3266     & 0.0590 & 3.81 & 2.1    \\
J0908.9-0938 &  137.2391  &  -9.6346   &    8.28 & Abell 0754     & 0.0530 & 2.96 & 1.8    \\

J1259.4+2757 &  194.8531  &  27.9523   &   19.95 & Coma Cluster   & 0.0230 & 4.32 & 5.1     \\
J1347.7-3253 &  206.9500  & -32.9000   &    5.05 & Abell 3571     & 0.0397 & 1.78 & 4.5     \\

J1511.0+0544 &  227.7500  &   5.7485   &    5.33 & Abell 2029     & 0.0770 & 2.71 & 0.8    \\
J1558.5+2714 &  239.6256  &  27.2417   &    7.11 & Abell 2142     & 0.0890 & 3.62 & 3.3     \\
J1638.8-6424 &  249.7136  & -64.4000   &    6.90 & Triangulum A.  & 0.0510  & 1.77 & 4.9  \\
J1712.3-2319 &  258.0914  & -23.3242   &   21.63 & Ophiucus       & 0.028 & 1.30  & 1.7  \\
J1920.9+4357 &  290.2405  &  43.9646   &   11.72 & Abell 2319     & 0.056 & 3.87 & 2.2     \\


\enddata


\end{deluxetable}

\begin{deluxetable}{lcccccc}
\tablewidth{0pt}
\tablecaption{Spectral parameters from combined XMM-Newton/XRT/Chandra and BAT fits (errors are 90\% C.L.) \label{tab:spec}}
\tablehead{
\colhead{NAME}    & \colhead{Flux\tablenotemark{a}} 
& \colhead{L$_\textrm{x}$\tablenotemark{a}} & \colhead{kT} & 
\colhead{$\Gamma$} & \colhead{model} 
& \colhead{$\chi^2$/dof}  \\
\colhead{}        & \colhead{\scriptsize (10$^{-11}$ cgs)}  
& \colhead{\scriptsize (10$^{43}$\,erg s$^{-1}$) }      & \colhead{ \scriptsize (keV)} & \\
}
\startdata

Perseus      & 3.90$^{+0.10}_{-1.65}$ & 2.7$^{+0.1}_{-1.1}$ 
&3.00$^{+0.40}_{-0.71}$/6.40$^{+0.62}_{-0.71}$ & 1.7$^{+0.3}_{-0.7}$
& apec+apec+pow & 152.8/144\\

Abell 3266 & 0.73$^{+0.10}_{-0.11}$ & 6.9$^{+0.9}_{-0.9}$ & 8.0$^{+0.4}_{-0.4}$ & & apec & 666.8/841  \\

Abell 0754 & 1.11$^{+0.04}_{-0.04}$ & 8.3$^{+0.3}_{-0.3}$ & 9.3$^{+0.4}_{-0.4}$ & & apec+pow & 1217.0/1072\\

Coma\tablenotemark{b}      & 2.33$^{+0.23}_{-0.22}$ & 3.0$^{+0.2}_{-0.4}$ & 
8.40$^{+0.25}_{-0.24}$/1.45$^{+0.21}_{-0.11}$ & & apec+apec & 846.5/856 \\
 
Abell 3571 & 0.63$^{+0.09}_{-0.06}$ & 2.7$^{+0.3}_{-0.4}$ & 6.0$^{+0.2}_{-0.2}$
& & apec & 723.9/1367 \\

Abell 2029 & 1.01$^{+0.16}_{-0.45}$ & 16.8$^{+2.4}_{-4.7}$ &  4.1$^{+1.7}_{-1.5}$/9.6$^{+2.0}_{-2.0}$ 
& & apec+apec & 394.2/363\\

Abell 2142 & 0.90$^{+0.10}_{-0.10}$ & 21.5$^{+3.5}_{-2.6}$ & 8.40$^{+0.64}_{-0.45}$ & & apec & 361.9/398\\

Triangulum A. & 1.30$^{+0.10}_{-0.10}$ & 8.8$^{+0.6}_{-0.2}$ & 9.30$^{+0.30}_{-0.30}$ & & apec & 925.8/1074\\

Ophiucus & 5.7$^{+0.5}_{-0.5}$ & 9.38$^{+0.28}_{-0.14}$ & 9.93$^{+0.24}_{-0.24}$
& & apec & 323.1/351\\

Abell 2319 & 1.56$^{+0.14}_{-0.14}$ & 13.0$^{+0.9}_{-0.8}$ & 9.23$^{+0.27}_{-0.27}$ & & apec & 
1151.3/1274 \\

\enddata

\tablenotetext{a}{Flux and Luminosities are computed in the 15--55\, keV band.}
\tablenotetext{b}{The spectral values reported for Coma are only representative for the 
source extraction region (i.e. 10\arcmin\ around the BAT centroid; see $\S$~\ref{subsec:coma}  for
more details).}


\end{deluxetable}

\begin{deluxetable}{lcccccccc}
\tablewidth{0pt}
\tablecaption{3$\sigma$ Upper limits on the non-thermal component and
Clusters' properties
\label{tab:ul}}
\tablehead{
\colhead{NAME}    & \colhead{CC\tablenotemark{a} ?} & \colhead{Merger ?}   &
\colhead{F$_{50-100\,keV}$\tablenotemark{b}}    & \colhead{B} & \colhead{S$_{\textrm{radio}}$} &
\colhead{$\nu_{\textrm{radio}}$} & $\alpha$ & \colhead{Ref\tablenotemark{c}} \\
\colhead{}   & \colhead{}  & \colhead{} & \colhead{\scriptsize (10$^{-12}$\,erg cm$^{-2}$ s$^{-1}$)}  
& \colhead{\scriptsize ($\mu$G)} & \colhead{\scriptsize (Jy)} & \colhead{\scriptsize (MHz)} & \colhead{}  & \colhead{}  \\
}
\startdata
Perseus	             &y&y&  \nodata      & \nodata     &  \nodata    &  \nodata   &  \nodata   & \nodata \\
Abell 3266           &n&y& $<$5.30 & $>$ 0.17  & 1.070 & 2700 & 0.95 & 1 \\
Abell 0754           &n&y& $<$6.50 & $>$ 0.10 & 0.086 & 1365 & 1.5 & 2 \\
Coma                 &n&y&  \nodata   & \nodata       &  \nodata   &  \nodata  &  \nodata   & \nodata \\ 
Abell 3571           &y$^{d}$&n$^{e}$& $<$11.5 & $>$ 0.03  & 0.0084 & 1380 & 1.5$^g$ & 3  \\
Abell 2029           &y$^{d}$&y& $<$4.83 & $>$ 0.25  & 0.528 & 1380  & 1.5$^g$ & 4    \\
Abell 2142           &y$^{f}$&y&$<$5.35 & $>$0.06  & 0.0183 & 1400 & 1.5$^g$ & 5   \\
Triangulum A. &y$^{f}$&y$^{f}$& $<$4.65 & $>$ 0.17  & $<$ 0.033 & 4850 &  1.5$^g$ & 6  \\
Ophiucus             &n&n& $<$5.89 & $>$ 0.11  & 6.4 & 160 & 2.0 & 7   \\ 
Abell 2319           &y$^{d}$&y& $<$3.41 & $>$ 0.10 & 1.0 & 610 & 0.92 & 8   \\
\enddata

\tablenotetext{a}{CC=Cool Core}
\tablenotetext{b}{BAT data alone were used to estimate the upper limits}
\tablenotetext{c}{References for the radio flux}
\tablenotetext{d}{Moderate CC}
\tablenotetext{e}{The morphology and temperature map indicate that it is a relaxed cluster, but the radio structure points at late stages of merging}
\tablenotetext{f}{Under discussion}
\tablenotetext{g}{Arbitrary spectral index}
\tablerefs{$1)$ \cite{bro91}, $2)$ \cite{fusco03}, $3)$  \cite{con98}, $4)$ \cite{gio00}, $5)$ \cite{condon93}, $6)$ \cite{sle77}, $7)$ \cite{fer97}.}

\end{deluxetable}

\begin{deluxetable}{lcc}
\tablewidth{0pt}

\tablecaption{Non-thermal emission from combined XMM-Newton/XRT/Chandra
 and BAT data.
\label{tab:ulxmm}}
\tablehead{
\colhead{NAME}    &
\colhead{F$_{50-100\,keV}$\tablenotemark{a}}    & 
\colhead{B\tablenotemark{b}} \\
\colhead{}   & \colhead{\scriptsize (10$^{-12}$\,erg cm$^2$ s$^{-1}$)}  
& \colhead{\scriptsize ($\mu$G)} \\
}
\startdata
Perseus	             &  \nodata   &  \nodata    \\
Abell 3266           &  $<$0.57   &  $>$ 0.55  \\
Abell 0754            &  \nodata   &  \nodata    \\
Coma                 &  \nodata   &  \nodata    \\
Abell 3571           &  1.4$^{+0.4}_{-0.4}$   &  $\sim$ 0.08 \\
Abell 2029           &  $<$1.27   &  $>$ 0.42  \\
Abell 2142           &  $<$1.50   &  $>$ 0.10  \\
Triangulum Australis &  $<$0.65   &  $>$ 0.39 \\
Ophiucus             &  $<$2.80   &  $>$ 0.15  \\
Abell 2319           &  $<$0.67   &  $>$ 0.15 \\
\enddata

\tablenotetext{a}{The flux has been estimated using a power-law spectrum
with a photon index of 2.0 in the 1--200\,keV energy band. Upper limits
are 99\,\% CL while errors are 90\,\% CL.}
\tablenotetext{b}{In order to compute the intensity of the magnetic
field we used the same Radio data reported in Tab.~\ref{tab:ul}}
\end{deluxetable}

\clearpage
\begin{table}[!ht]
\begin{center}
\caption{Comparison of different spectral fits to the clusters which show
a large deviation between the  ICM 
temperature as  measured below and above 10\,keV. As a reference for the
reader also the parameters of Coma are reported.
kT$_1$ and kT$_2$ are the temperatures of the two thermal models (in keV)
while norm. and $\Gamma$ are the normalization at 1\,keV 
(in ph cm$^{-2}$ s$^{-1}$ keV$^{-1}$) and the photon index of the
power-law model.
Frozen parameters do not have an error estimate.}
\label{tab:5}
\vspace{1cm}
\begin{tabular}{c|ccc}
Cluster     & Thermal & Thermal + power law & Thermal + Thermal \\
\hline
\hline
{\bf Abell 2029}   \\
kT$_1$       &   6.75$^{+0.52}_{-0.31}$  &6.78$^{+0.46}_{-0.33}$ & 4.1$^{+1.7}_{-1.5}$ \\
$\Gamma$     & & 2.0 & \\
norm.         &   &  1.55$^{+1.12}_{-1.15}\times 10^{-3}$& \\
kT$_2$       &   &  & 9.6$^{+2.0}_{-2.0}$  \\
$\chi^2$/dof &  407.3/364    & 402.1/363  & 394.2/363  \\
\hline
\hline
{\bf Triangulum A. }    \\
kT$_1$       &  9.30$^{+0.30}_{-0.30}$ & 9.25$^{+0.30}_{-0.28}$ &11.1$^{+0.34}_{-0.27}$\\
$\Gamma$     &  & 2.0\\
norm.         &  &$<1.40\times 10^{-4}$\\
kT$_2$       &  & & 1.63$^{+0.46}_{-0.27}$ \\
$\chi^2$/dof &  925.8/1074 & 925.8/1073 & 916.5/1072\\
\hline
\hline
{\bf Abell 2319}   &  \\
kT$_1$       &  9.23$^{+0.27}_{-0.27}$  & 9.33$^{+0.35}_{-0.52}$ & 11.2$^{+0.8}_{-1.0}$\\
$\Gamma$     & & 1.7$^{+0.2}_{-0.3}$\\
norm.         &  & 7.8$^{+2.7}_{-5.3}\times 10^{-4}$\\
kT$_2$       &  & & 1.9$^{+1.64}_{-0.40}$   \\
$\chi^2$/dof &  1151.34/1274   & 1139.9/1272  & 1127.8/1272  \\
\hline
\hline
{\bf Coma}   &  \\
kT$_1$       & 6.50$^{+0.09}_{-0.05}$ & 7.19$^{+0.16}_{-0.06}$ & 8.40$^{+0.25}_{-0.24}$ \\
$\Gamma$     &  & 2.11$^{+0.10}_{-0.13}$\\
norm         &  & 3.56$^{+0.46}_{-0.34}\times 10^{-3}$\\
kT$_2$       &  & & 1.45$^{+0.21}_{-0.11}$\\
$\chi^2$/dof &  1168.9/858 & 905.5/856 & 846.5/856\\
\hline
\hline


\end{tabular}
\end{center}
\end{table}

\end{document}